\documentclass[
 superscriptaddress,
 amsmath,
 amssymb,
 aps,
 pra,
 floatfix,
 twocolumn,
 longbibliography
]{revtex4-2}

\newcommand{\bra}[1]{\langle #1\rvert}
\newcommand{\ket}[1]{\lvert #1\rangle}
\newcommand{\ip}[2]{\bra{#1} #2\rangle}
\newcommand{\op}[2]{\ket{#1} \bra{#2}}
\newcommand{\exv}[1]{\langle #1\rangle}

\newcommand{\pd}[1]{\frac{\partial #1}{\partial t}}
\newcommand{\rpd}[1]{\partial_t #1}

\DeclareMathOperator{\Tr}{Tr}

\usepackage[varg]{newtxmath}
\usepackage{newtxtext}

\usepackage{graphicx}
\usepackage{dcolumn}
\usepackage{bm}
\usepackage{hyperref}
\usepackage{tikz}
\usetikzlibrary{calc}

\usepackage{orcidlink}

\usepackage{amsmath}
\usepackage{upgreek}
\usepackage{physics}
\usepackage{placeins}
\usepackage{comment}
\usepackage{mathrsfs}

\usepackage[plainruled,vlined]{algorithm2e}
\usepackage{ragged2e}
\makeatletter

\SetAlgoCaptionLayout{RevTeXAlgoCaptionLayout}
\makeatother

\begin{document}

\title{Nonclassical Many-Body Superradiant States with Interparticle and Spin-Momentum Entanglement}

\author{Jarrod T. Reilly\orcidlink{0000-0001-5410-089X}}
\affiliation{JILA and Department of Physics, University of Colorado, 440 UCB, Boulder, CO 80309, USA}
\author{Gage W. Harmon\orcidlink{0000-0001-6070-4470}}
\affiliation{Theoretische Physik, Universit\"at des Saarlandes, D-66123 Saarbr\"ucken, Germany}
\author{John Drew Wilson\orcidlink{0000-0001-6334-2460}}
\affiliation{JILA and Department of Physics, University of Colorado, 440 UCB, Boulder, CO 80309, USA}
\author{Murray J. Holland\orcidlink{0000-0002-3778-1352}}
\affiliation{JILA and Department of Physics, University of Colorado, 440 UCB, Boulder, CO 80309, USA}
\author{Simon B. J\"ager\orcidlink{0000-0002-2585-5246}}
\affiliation{Physikalisches Institut, University of Bonn, Nussallee 12, 53115 Bonn, Germany}

\date{\today}


\begin{abstract}
We present a cross-cavity system in which steady-state superradiance is achieved using solely collective dissipative dynamics.
Two cavities symmetrically couple an ensemble of four-level atoms by driving transitions between two electronic states and two motional states along perpendicular cavity axes.
Both cavities operate in the bad-cavity regime: one cavity mediates collective atomic decay, while the other cavity, together with a coherent drive, mediates collective pumping via an off-resonant Raman transition.
With this, we find steady-state superradiant states that possess nonclassical properties, such as super-Poissonian photon statistics. 
The system thus requires a beyond mean-field description, and so we develop an exact master equation simulation technique utilizing strong symmetries of the system's jump operators. 
Because superradiant decay is accompanied by a momentum impulse along the corresponding cavity axis, the system exhibits substantial hybrid entanglement between the atoms' spin and motional degrees of freedom at steady state.
We also demonstrate that heralded measurements of the two cavity outputs prepare a state with significant particle-particle entanglement with prospects for quantum-enhanced acceleration sensing.
\end{abstract}
 
{
\let\clearpage\relax
\maketitle
}

\section{Introduction}
Many-body quantum effects provide a rich platform for interdisciplinary physics, offering a host of useful nonclassical properties~\cite{Bloch,Amico,Weimer}.
One of the most studied many-body processes is Dicke superradiance~\cite{Dicke}, in which a collection of excited atoms or ions synchronize such that their dipoles constructively interfere to form a macroscopic dipole that collectively emits light at a rate exceeding that of independent emitters~\cite{Gross}.
Although conceptually simple, superradiance has a wide range of potential applications, including quantum sensing~\cite{Wang,Kucsko,Koppenhofer}, ultrastable lasers~\cite{Meiser,Reilly,Liu,Bychek,Dong}, dissipative generation of interparticle entanglement~\cite{PineiroOrioli,Lohof,Huang,Zhang,Hotter}, exploration of quantum effects on curved spacetime~\cite{Brito,Bekenstein,Cardoso}, probing the neutrino mass problem~\cite{Fukumi}, and even investigating quantum phenomena in the human brain~\cite{Celardo}.
While originally envisioned for closely spaced emitters in free space, superradiance also occurs in bad-cavity systems where photons escape the resonator before interacting appreciably with other atoms~\cite{Bonifacio,Bonifacio2}.
In the idealized case where two-level atoms are trapped at the same point in the cavity mode function, the system possesses permutation symmetry and can be described by a statistical mixture of unentangled coherent spin states~\cite{Wolfe,Rosario}. 
From a technical perspective, it remains unclear whether an approach based on such unentangled coherent spin states can be extended to more complex driven-dissipative multilevel superradiant systems. 
From a fundamental standpoint, this raises the question of which ingredients are genuinely required to generate true quantum entanglement. 
Addressing these issues will help establish a clear pathway toward quantum technologies that can be realized in state-of-the-art laboratory platforms.

Recently, there has been growing interest in using optical cavities to generate quantum correlations between the internal and external degrees of freedom of atoms~\cite{Gingrich,Jeske,Leonard,Mivehvar,Lang,Kroeze,Nishant,Kale,Ferri,Greve,ActuallyJarrodsPaper,Reilly2,Finger,Chelpanova,You,ActuallyJarrodsPaper3}, i.e., creating hybrid entanglement~\cite{VanLoock,Takeda,Andersen,Ulanov,Guccione,Shukla,Xu3,Li} between spin and momentum.
Generating and maintaining spin-momentum correlations is central to many phenomena, such as realizing topological quantum phases for quantum computing and information processing~\cite{Hasan}.
However, a major challenge is that composite systems of internal and external modes inherit noise from both degrees of freedom, making them highly susceptible to decoherence.
This motivates the need for fast, efficient, and robust state-preparation schemes.

To mitigate decoherence, one can engineer collective dissipation channels that stabilize the quantum system in a nonclassical many-body state~\cite{DallaTorre,Shankar,Reilly4,Ariunbold,Finger,Chelpanova}.
In such schemes, nonclassical states form fixed points of the dynamics, providing robustness against external perturbations.
The resulting many-body states can be dissipatively prepared and dynamically controlled, enabling the fast and stable generation of entangled states for quantum metrology or quantum information applications.
Many theoretical approaches pursuing this objective rely, either explicitly or implicitly, on mean-field approximations. 
However, such treatments fundamentally limit the amount and type of entanglement that can be described and quantified. 
A complete and efficient quantum description is therefore required to identify genuinely beyond-mean-field effects and to fully uncover the quantum potential of dissipatively generated many-body states.

In this work, we present a collective dissipative model whose steady-state properties cannot be captured accurately within mean-field theory. 
Instead, we develop a fully quantum description that takes advantage of the underlying symmetries of the collective spin model. 
Specifically, we introduce a cross-cavity setup that generates both interparticle entanglement and spin–momentum entanglement through fully collective steady-state superradiance.
The two cavities mediate electronic transitions between the atoms' internal states, while energy constraints restrict the dynamics to two momentum states along the respective cavity axes, as in Ref.~\cite{ActuallyJarrodsPaper}.
Both cavities operate in the bad-cavity regime, allowing their fields to be adiabatically eliminated, resulting in an atom-only master equation with collective jump operators.
Specifically, one cavity mediates collective decay, while the other cavity, together with a classical drive, mediates collective pumping via an off-resonant Raman transition, similar to Ref.~\cite{Reilly}. 
The two jump operators possess strong symmetries~\cite{Buca,Lieu,Nairn} which allows for exact diagonalization of the system's Liouvillian for $N \gg 10$ atoms when given a certain initial state. 
With our exact simulation, we demonstrate that the superradiant light output from one cavity transitions from nearly coherent to highly bunched when the collective pump rate exceeds the collective decay rate, with the opposite behavior for the other cavity.
This superradiant bunching behavior is a result of higher-order correlations that cannot be modeled effectively using mean-field techniques~\cite{Shankar,Shankar2}. 
Moreover, the momentum impulses associated with superradiant emission generate substantial hybrid entanglement between the spin and motional degrees of freedom along the cavity axes, even in a fully dissipative system. 
This again requires beyond-mean-field simulations to accurately model these nonclassical many-body states.
Finally, we show that heralded measurements of the emitted light prepare the system in a highly entangled state with potential applications in quantum-enhanced acceleration sensing.

The structure of the paper is as follows.
In Sec.~\ref{Sec:CrossCavity}, we introduce the dissipative model at the center of our analysis and describe our exact quantum master equation simulation techniques.
In Sec.~\ref{Sec:Superradiance}, we discuss the properties of the emitted supperadiant light fields.
Sec.~\ref{Sec:SpinMomentumEntanglement} presents the coherent information between spin and momentum, highlighting the system's hybrid entanglement.
In Sec.~\ref{Sec:Metrology}, we study the potential use of the system, under heralded measurements, for quantum-enhanced acceleration sensing.
Finally, Sec.~\ref{Sec:Conclusion} provides concluding remarks and an outlook for future work.

\section{Cross-Cavity System} \label{Sec:CrossCavity}

\subsection{Model} \label{Sec:Model}
\begin{figure}
    \begin{center}
    \includegraphics[width=\linewidth]{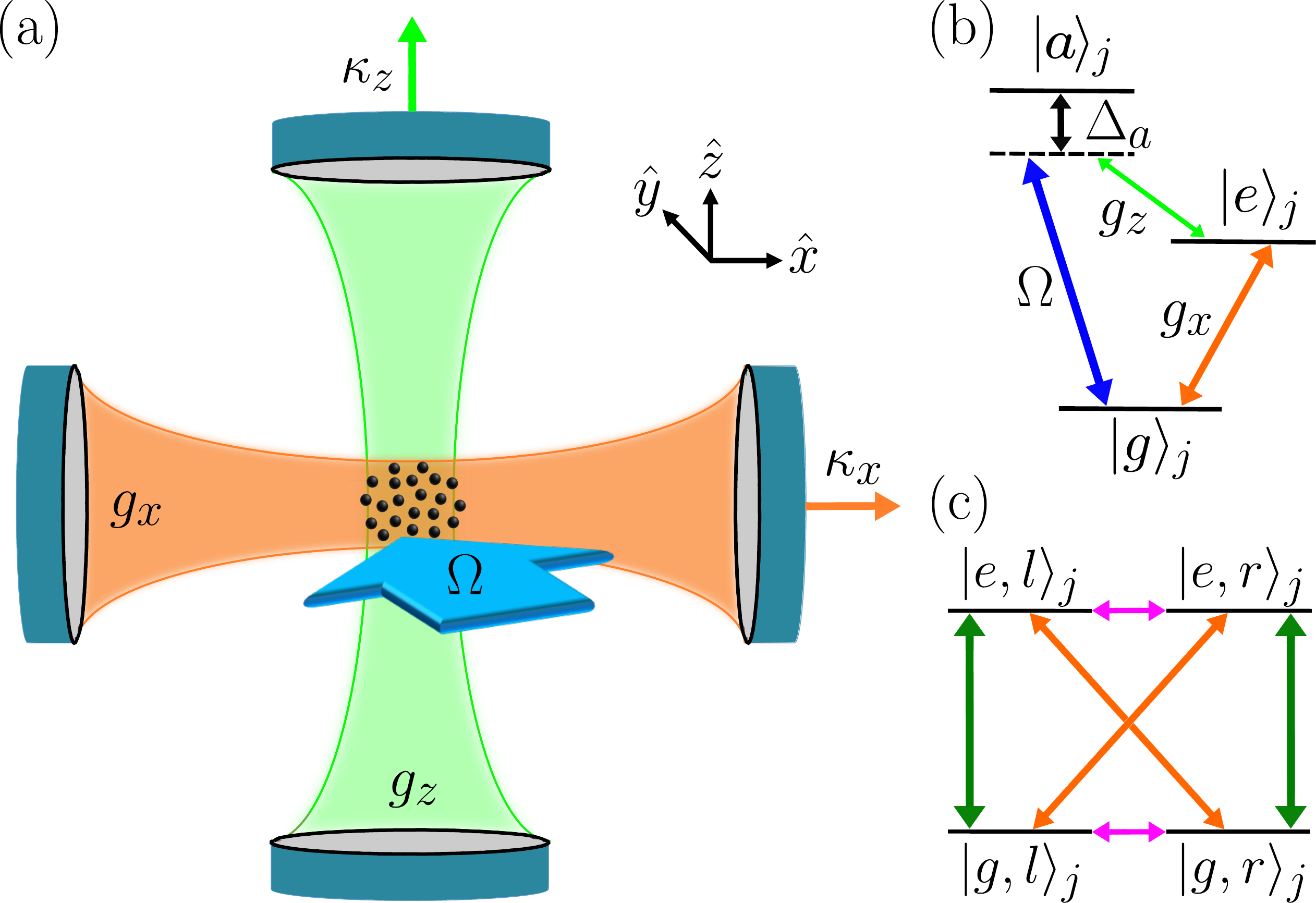}
    \end{center}
    \caption{(a) Schematic of the cross-cavity system, also pumped from the side by a coherent field. Both cavity fields can decay into free-space modes, while spontaneous emission from the atoms is neglected.
    (b) Level diagram of atom $j$ with three internal states, $\ket{g}_j$, $\ket{e}_j$, and $\ket{a}_j$.
    (c) Effective $\mathrm{SU}(4)$ system for atom $j$ constructed from the spin-$x$-momentum states of Eq.~\eqref{SU4_States}.
    Green arrows indicate $\hat{J}_{\pm}$, orange arrows indicate $\hat{E}_{\pm}$, and pink arrows indicate $\hat{K}_{\pm}$.}
    \label{Schematic}
\end{figure}
We consider the dual-cavity setup schematically shown in Fig.~\ref{Schematic}(a).
A cloud of $N$ atoms is coupled to two perpendicular single-mode cavity fields oriented along the $x$- and $z$-directions, with coupling constants $g_x$ and $g_z$.
The cavity fields have mode profiles $\cos(k_x \hat{x})$ and $\cos(k_z \hat{z})$ with wavenumbers $k_x$ and $k_z$, where $\hat{x}$ and $\hat{z}$ are the position operators.
The atomic internal structure consists of a ground state $\ket{g}$ and two excited states $\ket{e}$ and $\ket{a}$.
As shown in Fig.~\ref{Schematic}(b), the $x$-oriented cavity couples resonantly to the $\ket{g} \leftrightarrow \ket{e}$ transition.
The atoms are coherently pumped on the $\ket{g} \leftrightarrow \ket{a}$ transition by a $y$-oriented laser with Rabi frequency $\Omega$ and detuning~$\Delta_a$.
The $z$-oriented cavity couples off-resonantly, with the same detuning $\Delta_a$, to the $\ket{e} \leftrightarrow \ket{a}$ transition.
In the large-detuning limit, $\abs{\Delta_a} \gg \sqrt{N} g_z, \abs{\Omega}$, the auxiliary state $\ket{a}$ can be adiabatically eliminated, resulting in a resonant two-photon process mediated by the laser and a $z$-cavity photon~\cite{Reiter,Reilly}.
Finally, both cavities decay into free-space electromagnetic modes at rates $\kappa_x$ and $\kappa_z$.

We assume all atoms are initialized in the state $\ket{g, -p_x, -p_z}$, with momentum $p_i = \hbar k_i / 2$ for $i = x,z$.
We work in a parameter regime similar to Ref.~\cite{ActuallyJarrodsPaper}, where the momentum states $\ket{\pm p_x}$ and $\ket{\pm p_z}$ are decoupled from the rest of the quadratic kinetic-energy spectrum (see Appendix~\ref{Appendix:MomentumReduction} for details).
This allows us to restrict the dynamics to the states
\begin{equation}
    \{ \ket{g, p_x, p_z}, \ket{g, - p_x, - p_z}, \ket{e, p_x, - p_z}, \ket{e, - p_x, p_z} \}
\end{equation}
of each atom, all having the same kinetic energy.
Since the product of the $x$- and $z$-momenta is positive for atoms in the ground state and negative for atoms in the excited state, the $z$-momentum can be inferred from the internal state and the $x$-momentum.
We can therefore omit the $z$-momentum labels as they are redundant, and describe the dynamics using the states
\begin{equation} \label{SU4_States}
    \left\{ \ket{g, r}, \ket{g, l}, \ket{e, r}, \ket{e, l} \right\}, 
\end{equation}
where $l$ and $r$ indicate whether the atom is moving left (negative $x$; $\ket{l} = \ket{-\hbar k_x / 2}$) or right (positive $x$; $\ket{r} = \ket{\hbar k_x / 2}$) along the $x$-axis. 

We now assume both cavities operate in the bad-cavity regime, where the linewidth of the $z$-oriented cavity exceeds the collective pumping rate, $N W = N g_z^2 \Omega^2 / (4 \Delta_a^2 \kappa_z) \ll \kappa_z$, and the linewidth of the $x$-oriented cavity exceeds the collective decay rate, $N \Gamma_c = N g_x^2 / \kappa_x \ll \kappa_x$.
In the regime where the decay of both cavities are much faster than the collective decoherence $N W, N \Gamma_c \ll \kappa_x, \kappa_z$, as well as any single-particle decoherence, both cavity modes can be adiabatically eliminated over a coarse-grained timescale yielding a simplified atom-only master equation~\cite{Jager}.
Accordingly, we obtain an effective Born-Markov master equation for the atomic density matrix $\hat{\rho}$:
\begin{equation} \label{AtomicMasterEq}
    \hat{\mathcal{L}} \hat{\rho} \equiv \pd{\hat{\rho}} = \hat{\mathcal{D}} \left[ \sqrt{W} \hat{J}_+ \right] \hat{\rho} + \hat{\mathcal{D}} \left[ \sqrt{\Gamma_c} \hat{E}_- \right] \hat{\rho},
\end{equation}
with the Liouvillian $\hat{\mathcal{L}}$ and Lindblad superoperator 
\begin{equation}
    \hat{\mathcal{D}} [\hat{O}] \hat{\rho} = \hat{O} \hat{\rho} \hat{O}^{\dagger} - \frac{1}{2} \left( \hat{O}^{\dagger} \hat{O} \hat{\rho} + \hat{\rho} \hat{O}^{\dagger} \hat{O} \right). 
\end{equation}
The full derivation of Eq.~\eqref{AtomicMasterEq} is presented in Appendix~\ref{Appendix:ModelDerivation}.
In Eq.~\eqref{AtomicMasterEq}, we introduced the collective operators $\hat{J}_+ = \sum_{j = 1}^N \op{e}{g}_j$ and $\hat{E}_- = \sum_{j = 1}^N \op{g}{e}_j \otimes \hat{s}^x_j$, with $\hat{s}_j^x = \op{l}{r}_j + \op{r}{l}_j$.
We also define the collective $x$-momentum operator $\hat{K}_- = \sum_{j = 1}^N \op{l}{r}_j$.
The actions of these collective operators are illustrated in Fig.~\ref{Schematic}(c).
For each dipole operator, $\hat{O} \in \{ \hat{J}, \hat{K}, \hat{E} \}$, we can construct the generators of the corresponding $\mathrm{SU}(2)$ subgroups,
\begin{equation} \label{SU2Generators}
    \hat{O}_x = \frac{\hat{O}_+ + \hat{O}_-}{2}, \; \hat{O}_y = \frac{i \hat{O}_- - i \hat{O}_+}{2}, \; \hat{O}_z = \frac{[\hat{O}_+, \hat{O}_-]}{2},
\end{equation}
as well as the quadratic Casimir operator,
\begin{equation} \label{CasimirOp}
    \hat{O}^2 = \hat{O}_x^2 + \hat{O}_y^2 + \hat{O}_z^2.
\end{equation} 
We note in passing that $\hat{E}_z = \hat{J}_z$~\cite{ActuallyJarrodsPaper}.
We have also neglected single-particle spontaneous emission from $\ket{e}_j$ and $\ket{a}_j$ in Eq.~\eqref{AtomicMasterEq}, which is justified when the collective rates $N W$ and $N \Gamma_c$ far exceed any single-particle decay rates.
The master equation~\eqref{AtomicMasterEq} lies at the center of our theoretical analysis and describes the dissipative dynamics of many atoms with coupled internal and external degrees of freedom.

\subsection{Simulation Techniques} \label{Sec:SimTechniques}
For notational simplicity, we assume that $N$ is even throughout, as the extension to odd $N$ is straightforward (the smallest $\hat{O}^2$ eigenvalue label is $1/2$ instead of $0$).
In general, the Hilbert space dimension scales as $4^N$, so the Liouville space in Eq.~\eqref{AtomicMasterEq} scales as $16^N$, making exact numerical simulations feasible only for very small atom numbers.
To enable efficient simulations, we exploit the permutation symmetry of the master equation and the collective nature of the Liouvillian [which commutes with the $\mathrm{SU}(4)$ Casimir operators] to restrict the dynamics to the $\mathrm{SU}(4)$ bosonic subspace, describing collective atomic states via Schwinger bosons of the states in Eq.~\eqref{SU4_States}~\cite{Manu,ActuallyJarrodsPaper,ActuallyJarrodsPaper3}.
In this irreducible representation, state vectors scale as $(N+1)(N+2)(N+3)/6 \sim \mathcal{O}(N^3)$~\cite{Xu}, so the Liouville space scales as $\mathcal{O}(N^6)$.
This simulation basis is used in Secs.~\ref{Sec:SpinMomentumEntanglement} and~\ref{Sec:Metrology}.

On top of this $\mathrm{SU}(4)$ structure, the system exhibits an additional strong symmetry arising from the number of atoms, $N_{\pm}$, in the states
\begin{equation}
    \ket{\pm} = \frac{\ket{r} \pm \ket{l}}{\sqrt{2}},
\end{equation}
being conserved by both of the system's jump operators. 
Therefore, the system can be viewed as two spin species 
\begin{equation}
    \hat{M}_{\pm} = \frac{\hat{J}_{\pm} - \hat{E}_{\pm}}{2}, \quad \hat{P}_{\pm} = \frac{\hat{J}_{\pm} + \hat{E}_{\pm}}{2},
\end{equation}
which couple with different phases to the light fields~\cite{Nairn}.
From these lowering and raising operators, we can again construct $\mathrm{SU}(2)$ subgroups using Eqs.~\eqref{SU2Generators} and~\eqref{CasimirOp}.
From this, the strong symmetry condition is given by commutators with the Casimir operators $\hat{M}^2$ and $\hat{P}^2$, $[\hat{M}^2, \hat{J}_+] = [\hat{M}^2, \hat{E}_-] = 0$ and $[\hat{P}^2, \hat{J}_+] = [\hat{P}^2, \hat{E}_-] = 0$~\cite{Buca,Lieu}. 
We then find that the dipole lengths $M$ and $P$, given by the eigenvalues of the Casimir operators,
\begin{equation}
    \hat{M}^2 \ket{M} = M (M + 1) \ket{M}, \quad \hat{P}^2 \ket{P} = P (P + 1) \ket{P},
\end{equation}
satisfy the normalization condition $M + P = N/2$.
Thus, both can be expressed in terms of a single quantum number $\ell = N/2 + P - M$, which ranges over $\ell \in \{0, 1, \ldots, N\}$, with $M = (N - \ell)/2$ and $P = \ell/2$.
Using this, along with the eigenvalues $m$ and $p$ of the operators $\hat{M}_z$ and $\hat{P}_z$,
\begin{equation}
    \hat{M}_z \ket{m} = m \ket{m}, \quad \hat{P}_z \ket{p} = p \ket{p},
\end{equation}
we can transform to a basis $\{ \ket{\ell,m,p}, \, \forall \ell, m, p \}$, where $m \in \{-(N - \ell)/2, -(N - \ell)/2+1, \ldots, (N - \ell)/2\}$ and $p \in \{-\ell/2, -\ell/2+1, \ldots, \ell/2\}$.
In this basis, the collective operators for the $J$ and $E$ subgroups become block-diagonal,
\begin{equation}
    \hat{J}_{\pm} = \bigoplus_{\ell = 0}^N \hat{J}_{\pm}^{(\ell)}, \quad \hat{E}_{\pm} = \bigoplus_{\ell = 0}^N \hat{E}_{\pm}^{(\ell)},
\end{equation}
with no off-diagonal elements between different $\ell$ layers, $\bra{\ell,m,p} \hat{O} \ket{\ell',m',p'} \propto \delta_{\ell,\ell'}$.
Similarly, the generators in Eq.~\eqref{SU2Generators} and the Casimir operator in Eq.~\eqref{CasimirOp} for the $J$ and $E$ subgroups are block-diagonal.
Consequently, the Liouvillian superoperator $\hat{\mathcal{L}}$ is also block-diagonal, so the dynamics in different blocks are decoupled and Eq.~\eqref{AtomicMasterEq} can be solved block by block, greatly reducing the numerical complexity of our simulations.
To see this, one writes
\begin{equation} \label{BlockDiagMatrices}
    \begin{aligned}
\left( A_1 \oplus D_1 \right) M_2 &= \left( 
\begin{array}{c|c}
    A_1 A_2 & A_1 B_2 \\ 
    \hline
    D_1 C_2 & D_1 D_2 \\
\end{array} 
\right), \\
M_1 \left( A_2 \oplus D_2 \right) &= \left( 
\begin{array}{c|c}
    A_1 A_2 & B_1 D_2 \\ 
    \hline
    C_1 A_2 & D_1 D_2 \\
\end{array} 
\right),
    \end{aligned}
\end{equation}
with
\begin{equation}
    M_i = \left( 
\begin{array}{c|c}
    A_i & B_i \\ 
    \hline
    C_i & D_i \\
\end{array} 
\right),
\end{equation}
where $A_i,B_i,C_i,D_i$ are submatrices of $M_i$ ($i \in \{ 1, 2 \}$). 
Importantly, for a block diagonal operator $\hat{O} = \bigoplus_{\ell} \hat{O}_{\ell}$, we have
\begin{equation} \label{BlockDiagonalExp}
    \exv{\hat{O}} = \Tr[\hat{O} \hat{\rho}] = \sum_{\ell = 0}^N c_{\ell} \Tr[\hat{O}_{\ell} \hat{\rho}_{\ell}],
\end{equation}
where we have decomposed the density matrix as block diagonal elements $\hat{\rho}_{\ell}$ and the coherences between blocks~$\hat{\rho}_c$,
\begin{equation} \label{rhoDecomp}
    \hat{\rho} = \bigoplus_{\ell = 0}^N c_{\ell} \hat{\rho}_{\ell} + \hat{\rho}_c.
\end{equation}
Here, $\hat{\rho}_{\ell}$ is a properly normalized density matrix in the $\ell$-subspace and the coefficient $c_{\ell}$ represents its probability with $\sum_{\ell} c_{\ell} = 1$. 
Due to the strong symmetry of the jump operators, the coefficients are time-independent $\rpd{c_{\ell}} = 0$.
Therefore, if we are only interested in expectation values that consist of sums and products of $J$ and $E$ operators, we only need to evolve the master equation of each block,
\begin{equation}
    \hat{\mathcal{L}}_{\ell} \hat{\rho}_{\ell} \equiv \pd{\hat{\rho}_{\ell}} = \hat{\mathcal{D}} \left[ \sqrt{W} \hat{J}_+^{(\ell)} \right] \hat{\rho}_{\ell} + \hat{\mathcal{D}} \left[ \sqrt{\Gamma_c} \hat{E}_-^{(\ell)} \right] \hat{\rho}_{\ell},
\end{equation}
and then calculate the sum in Eq.~\eqref{BlockDiagonalExp}.
The largest blocks scale as $\mathcal{O}(N^4)$, and so leads to a large reduction of the simulation complexity for large atom numbers $N \gg 1$. 
We will use the $\{ \ket{\ell,m,p} \}$ basis throughout Sec.~\ref{Sec:Superradiance}.

\section{Properties of the Superradiant Light Fields} \label{Sec:Superradiance}
Having established the tools to efficiently simulate the master equation [Eq.~\eqref{AtomicMasterEq}], we now use them to examine the light output of the two cavity modes.
In the bad-cavity regime, the cavity modes are effectively slaved to the respective dipoles, $\hat{a}_x \sim \hat{E}_-$ and $\hat{a}_z \sim \hat{J}_+$ [see Eq.~\eqref{aSlaved}].
The photon fluxes from the respective cavities therefore become $\Gamma_c \exv{\hat{E}_+ \hat{E}_-}$ and $W \exv{\hat{J}_- \hat{J}_+}$.
Superradiance is identified by an $N^2$ scaling of the photon flux, resulting from the constructive interference of individual dipoles into a macroscopic dipole~\cite{Gross}.
To determine whether the system exhibits superradiance over a broad parameter regime at steady state, we calculate the kernel of the Liouvillian superoperator from Eq.~\eqref{AtomicMasterEq},
$\hat{\mathcal{L}} \hat{\rho}_{\mathrm{ss}} = 0$.
Since we are only interested in $J$ and $E$ observables, we can perform this calculation block by block, $\hat{\mathcal{L}}_{\ell} \hat{\rho}_{\ell,\mathrm{ss}} = 0$,
and then compute expectation values using Eq.~\eqref{BlockDiagonalExp}.
Due to the strong symmetries in the system, the steady-state coefficients remain the same as those of the chosen initial state, $\rpd{c_{\ell}} = 0$.
This is what determines the steady-state from a large degeneracy of zero eigenvalues of the Liouvillian~\cite{Zhang2}, and it is this ``memory'' of initial conditions that has garnered much interest in error correction and protection~\cite{Lieu,Santos,Caspel}.
As noted in Sec.~\ref{Sec:Model}, we assume the system begins in the state
\begin{widetext}
\begin{equation} \label{InitialState}
    \hat{\rho}_0 = \op{g,l}{g,l}^{\otimes N} = \frac{1}{2^N} \left( \bigoplus_{\ell = 0}^N \sqrt{\binom{N}{\ell}} \ket{\ell,- (N-\ell)/2,\ell/2} \right) \left( \bigoplus_{\ell' = 0}^N \sqrt{\binom{N}{\ell'}} \bra{\ell',- (N-\ell')/2,\ell'/2} \right),
\end{equation}
\end{widetext}
so that the coefficients of the $\ell$-blocks are
\begin{equation}
    c_{\ell} = \frac{1}{2^N} \binom{N}{\ell}. 
\end{equation}
We then use this to calculate steady-state expectation values of block-diagonal operators, $\exv{\oplus_{\ell} \hat{O}_{\ell}}_{\mathrm{ss}}$.

\begin{figure}
    \centerline{\includegraphics[width=\linewidth]{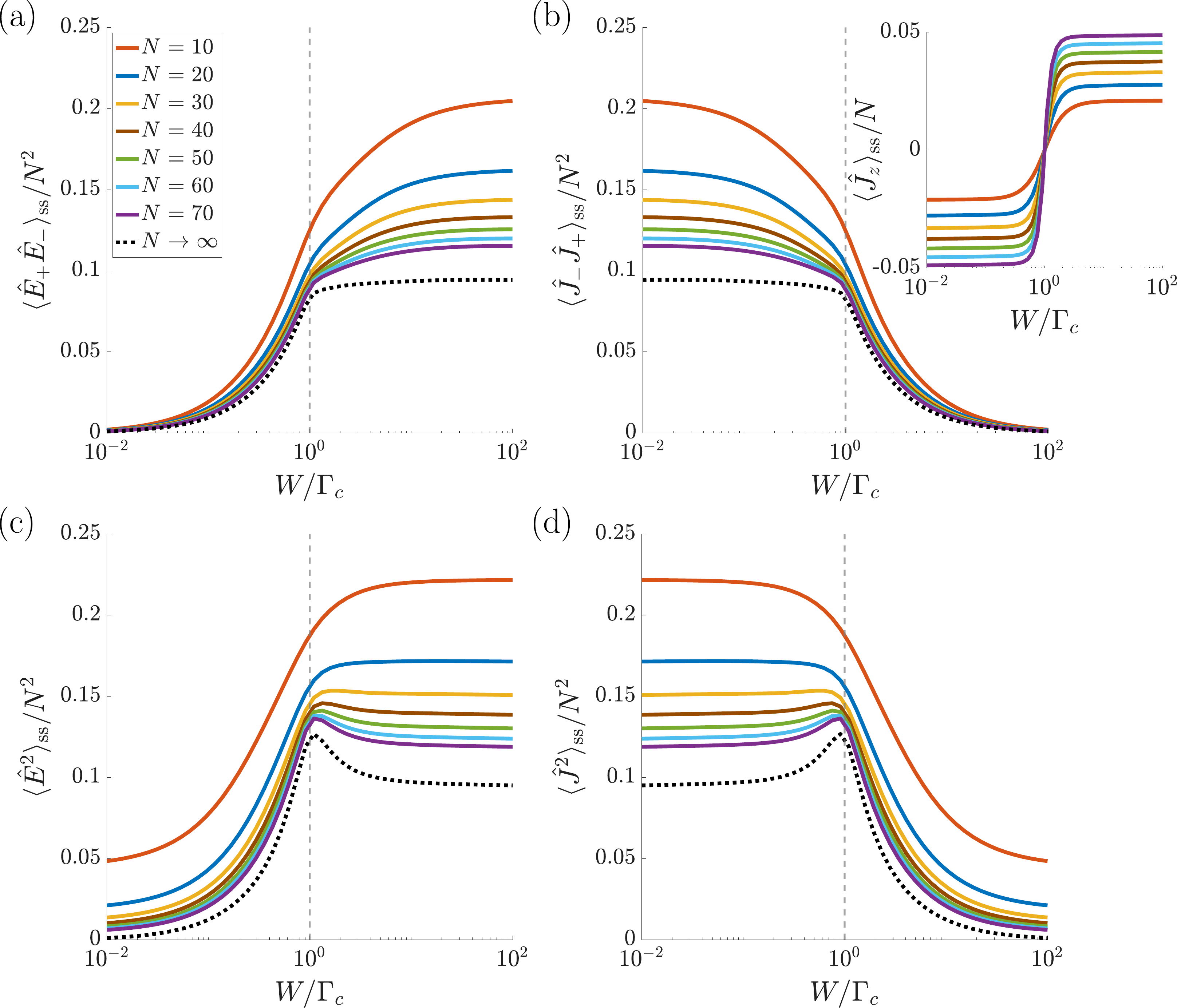}}
    \caption{Intensity of the (a) $x$-cavity and (b) $z$-cavity, signifying superradiance when $\mathcal{O}(N^2)$.
    The inset in (b) displays the spin inversion around the transition.
    Expectation values of the quadratic Casimir operator for the (c) $E$ and (d) $J$ subgroups.
    The $N \rightarrow \infty$ curves (dotted black lines) are from the fit in Eq.~\eqref{fN_fit}, suggesting a potential phase transition at $W = \Gamma_c$.}
    \label{LightOutput}
\end{figure}
In Figs.~\ref{LightOutput}(a) and~(b), we show the output intensities from the $x$- and $z$-cavities, $\exv{\hat{E}_+ \hat{E}_-}_{\mathrm{ss}}$ and $\exv{\hat{J}_- \hat{J}_+}_{\mathrm{ss}}$.
At steady state, the two intensities are related by
\begin{equation} \label{IntSS}
    \Gamma_c \exv{\hat{E}_+ \hat{E}_-}_{\mathrm{ss}} = W \exv{\hat{J}_- \hat{J}_+}_{\mathrm{ss}},
\end{equation}
which follows from the expectation value of the Heisenberg-Langevin equation for $\hat{J}_z$,
\begin{equation}
    \pd{} \exv{\hat{J}_z} = W \exv{\hat{J}_- \hat{J}_+} - \Gamma_c \exv{\hat{E}_+ \hat{E}_-}.
\end{equation}
Both cavities' intensities exhibit a significant $N^2$ scaling, signaling superradiant emission at steady state, but with different magnitude in different parameter regimes: the $x$-cavity is strongly superradiant when $W > \Gamma_c$, while the $z$-cavity is strongly superradiant when $W < \Gamma_c$.
In the opposite regime, $W < \Gamma_c$ for the $x$-cavity and $W > \Gamma_c$ for the $z$-cavity, the superradiant ($N^2$) component of the light field gradually shrinks towards zero as $\abs{\log_{10}( W/ \Gamma_c)} \rightarrow 2$.
This arises from the steady-state inversion of the internal states, $\exv{\hat{J}_z}_{\mathrm{ss}}$, switching from negative to positive as the collective pumping rate surpasses the collective decay rate, as shown in the inset of Fig.~\ref{LightOutput}(b). 
Unlike the fully collective $\mathrm{SU}(2)$ system of Ref.~\cite{Haake}, our model exhibits steady-state superradiant emission away from $W \approx \Gamma_c$ due to its multi-level nature~\cite{Reilly,Yu}.
Figures~\ref{LightOutput}(c) and~(d) show the steady-state expectation values of the quadratic Casimir operators, $\exv{\hat{E}^2}_{\mathrm{ss}}$ and $\exv{\hat{J}^2}_{\mathrm{ss}}$, which are approximately given by the square of the respective dipole lengths.
The $\mathrm{SU}(4)$ structure allows these dipole lengths to vary from their maximum value $N/2$, unlike the fully collective $\mathrm{SU}(2)$ case.
In the respective superradiant regimes ($W > \Gamma_c$ for $E$, $W < \Gamma_c$ for $J$), the steady-state dipole lengths remain roughly constant, while in the opposite regime they quickly shrink toward zero.
Interestingly, this allows one cavity to be subradiant while the other is superradiant. In the extreme, one dipole can have high overlap with a supersinglet~\cite{IloOkeke} ($J=0$ or $E=0$), while the other retains a large macroscopic component.
The supersinglet is a dark state for its cavity, satisfying $\hat{E}_{\pm} \ket{E=0} = \hat{J}_{\pm} \ket{J=0} = 0$. 

We now consider the thermodynamic limit $N \rightarrow \infty$.
For this, we fit the data to
\begin{equation} \label{fN_fit}
    f(N) = X + \frac{Y}{N} + \frac{Z}{N^2}.
\end{equation}
The dotted black lines in Fig.~\ref{LightOutput} show the thermodynamic limit $X(W)$ from fits of $f(N) = \exv{\hat{O}}_{\mathrm{ss}} / N^2$, with $\hat{O} \in \{ \hat{E}_+ \hat{E}_-, \hat{J}_- \hat{J}_+, \hat{E}^2, \hat{J}^2 \}$.
We see that in the superradiant regime of the respective cavities, the intensity has a significant superradiant component and the respective Casimir operator's expectation value is macroscopic. 
In the opposite regime, both the intensity and the expectation value of the respective Casimir operator gradually shrink towards zero when $\abs{\log_{10}( W/ \Gamma_c)} \rightarrow 2$ as the respective dipole heads towards its supersinglet state.
From these fits, we find evidence of a non-analytic change in the light output from either cavity at $W = \Gamma_c$.
This is possibly a signature of a fully dissipative phase transition in the system.

\begin{figure}
    \centerline{\includegraphics[width=\linewidth]{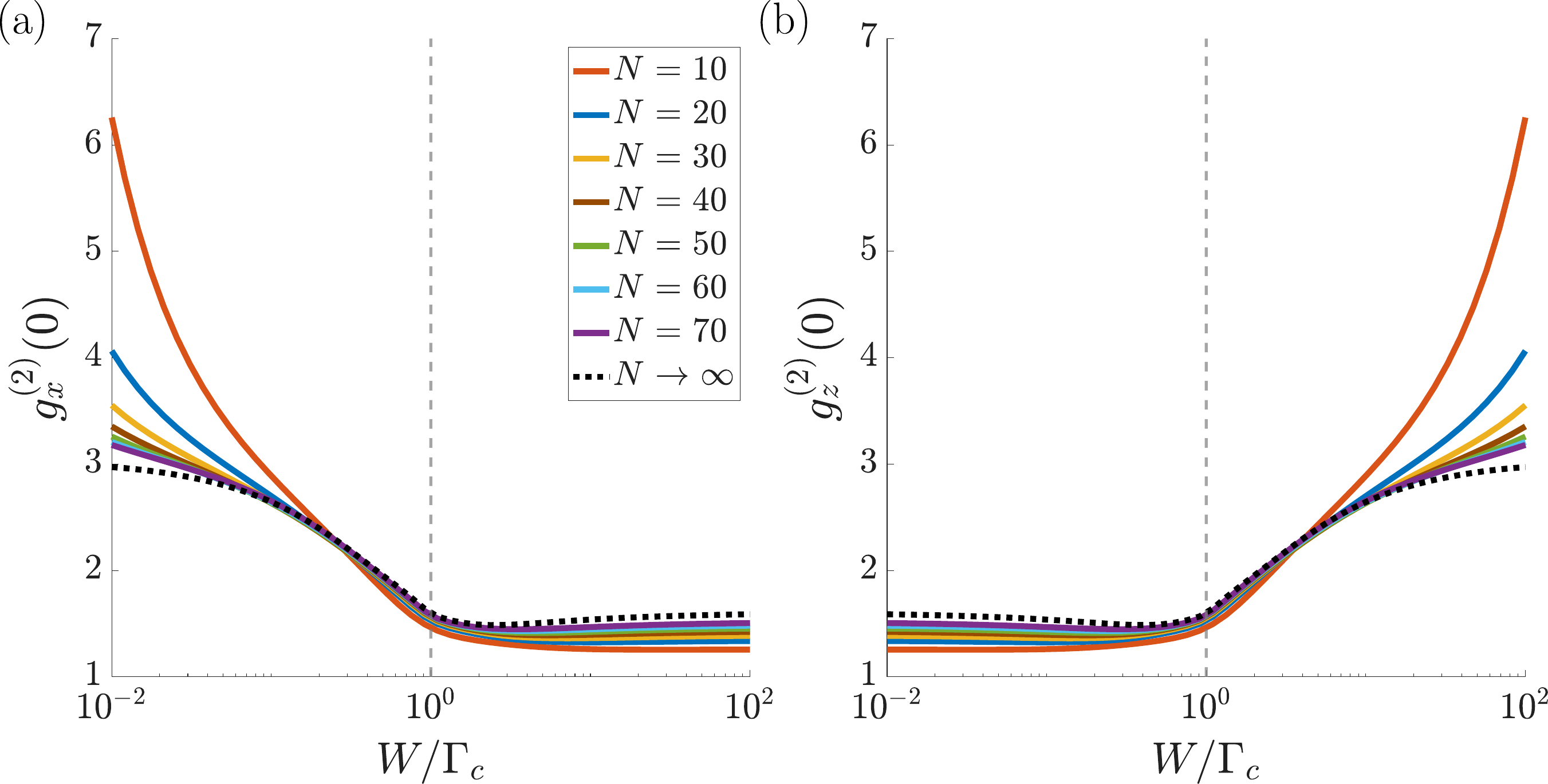}}
    \caption{Second-order coherence at zero time delay for the (a) $x$-cavity and (b) $z$-cavity. 
    The $N \rightarrow \infty$ curves (dotted black lines) are from the fit in Eq.~\eqref{fN_fit}.}
    \label{LightProperties}
\end{figure}
It may be surprising that the system remains superradiant in the $x$-cavity when $W \gg \Gamma_c$, rather than saturating the transition as in Refs.~\cite{Meiser,Meiser2,Meiser3,Reilly}, and similarly for the $z$-cavity when $\Gamma_c \gg W$.
One possible explanation, inspired by the supersinglet discussion above, is that the steady state has high overlap with collective dark states of one cavity~\cite{PineiroOrioli}, so large superradiant bursts are interspersed with periods of no photon emission which leads to extreme photon bunching (as outlined below).
This likely arises from destructive interference between the two channels $\ket{g,l} \rightarrow \ket{e,l} \rightarrow \ket{g,r}$ and $\ket{g,r} \rightarrow \ket{e,r} \rightarrow \ket{g,l}$.
High overlap with dark states also explains why the inversion in the inset of Fig.~\ref{LightOutput}(b) remains nearly constant for $W \lesssim \Gamma_c$ or $W \gtrsim \Gamma_c$, even when the pump rate changes by two orders of magnitude~\footnote{The steady-state inversion in the inset of Fig.~\ref{LightOutput}(b) can be compared to the inversion for the collective $\mathrm{SU}(2)$ model~\cite{Haake} $\rpd \hat{\rho}_2 = \hat{\mathcal{D}}[\sqrt{W} \hat{J}_+] \hat{\rho}_2 + \hat{\mathcal{D}}[\sqrt{\Gamma_c} \hat{J}_-] \hat{\rho}_2$ where the inversion quickly goes to $\pm N / 2$ away from $W \approx \Gamma_c$.}.
For example, in the $W \gg \Gamma_c$ region where $\exv{\hat{J}^2}_{\mathrm{ss}} \sim 0$, the state represented in the Dicke triangle for $J$ is concentrated near the vertex at $J = 0$ [similar to Fig.~1(d) of Ref.~\cite{Shankar}].
Since $J\ll N/2$ in the highly antisymmetric irreducible representations of the $J$ $\mathrm{SU}(2)$ subgroup, and the eigenvalues $m$ of $\hat{J}_z$ are restricted to $-J \leq m \leq J$, the inversion remains near zero even when pumping and decay rates differ by orders of magnitude.
In other words, there are $(N/2 - J)$ many singlet pairs in the $J$ $\mathrm{SU} (2)$ subgroup [and, due to the total $\mathrm{SU} (4)$ symmetry, the same number of singlet pairs in the $K$ $\mathrm{SU} (2)$ subgroup~\cite{ActuallyJarrodsPaper3}] whose inversion is zero, and so the collective inversion can only come from the remaining symmetric state between $2 J \ll N$ atoms~\footnote{One can see this directly from the method of Young tableaux~\cite{Georgi,Pfeifer,Eichmann}.}.
Because $\hat{E}_z = \hat{J}_z$, the highly populated states in the corresponding Dicke triangle for the $E$ $\mathrm{SU}(2)$ subgroup also remain near the center vertically (near the equator of the collective Bloch sphere), even though the system has a macroscopic dipole length and is superradiant.
The same argument applies when $\exv{\hat{E}^2}_{\mathrm{ss}} \sim 0$ for $W \ll \Gamma_c$.
Near the critical point $W = \Gamma_c$, both $\exv{\hat{J}^2}_{\mathrm{ss}}$ and $\exv{\hat{E}^2}_{\mathrm{ss}}$ reach their maximum in the large-$N$ limit, suggesting the state is highly symmetric with respect to both $J$ and $E$ subgroups, so the dark subspaces of each cavity are not heavily populated.
In Secs.~\ref{Sec:SpinMomentumEntanglement} and~\ref{Sec:Metrology}, we show how this distinction leads to significant differences in the state's utility for quantum information and quantum sensing in the regimes $W \gg \Gamma_c$ and $W \ll \Gamma_c$ compared to $W \approx \Gamma_c$.

To further investigate whether dark states are populated, we study the intensity fluctuations of each cavity.
These fluctuations can be measured using a Hanbury-Brown-Twiss–type setup~\cite{Mandel}, as in Ref.~\cite{Meiser3}.
We therefore introduce the second-order coherence functions of the two cavity fields,\begin{equation}
    \begin{aligned}
g_x^{(2)} (\tau) &= \frac{\exv{\hat{E}_+ (0) \hat{E}_+ (\tau) \hat{E}_- (\tau) \hat{E}_- (0)}}{\exv{\hat{E}_+ (0) \hat{E}_- (0)} \exv{\hat{E}_+ (\tau) \hat{E}_- (\tau)}}, \\
g_z^{(2)} (\tau) &= \frac{\exv{\hat{J}_- (0) \hat{J}_- (\tau) \hat{J}_+ (\tau) \hat{J}_+ (0)}}{\exv{\hat{J}_- (0) \hat{J}_+ (0)} \exv{\hat{J}_- (\tau) \hat{J}_+ (\tau)}},
    \end{aligned}
\end{equation}
where $\tau = 0$ is taken to correspond to the system already being in steady state.
We are primarily interested in the zero–time-delay case, $\tau = 0$,
\begin{equation}
    \begin{aligned}
g_x^{(2)} (0) &= \frac{\exv{\hat{E}_+ \hat{E}_+ \hat{E}_- \hat{E}_-}_{\mathrm{ss}}}{\exv{\hat{E}_+ \hat{E}_-}_{\mathrm{ss}}^2}, \\
g_z^{(2)} (0) &= \frac{\exv{\hat{J}_- \hat{J}_- \hat{J}_+ \hat{J}_+}_{\mathrm{ss}}}{\exv{\hat{J}_- \hat{J}_+}_{\mathrm{ss}}^2},
    \end{aligned}
\end{equation}
which are related to the fluctuations of the output intensities $\hat{I}_j$ by~\cite{Meiser3}
\begin{equation}
    \Delta I_j^2 = I_j^2 \left[ g_j^{(2)} (0) - 1 \right] + B_j I_j.
\end{equation}
Here, $j \in \{x,z\}$, $I_j = \exv{\hat{I}_j}$, and $B_j$ is the bandwidth of the respective photodetector, which is assumed to be much larger than the cavity bandwidth.
We display $g_j^{(2)}(0)$ for the two cavities across the phase transition in Fig.~\ref{LightProperties}.
In both regimes, we find that both cavities exhibit super-Poissonian photon statistics, $g_j^{(2)}(0) > 1$.
This implies that photons emitted from either cavity arrive at the photodetector in bunches, which typically results from superradiant bursts~\cite{Shankar2,Temnov}. 
When the respective cavities are in the regime where the $N^2$ component of their intensity shrinks toward zero, we find that they do not simply thermalize [$g_j^{(2)}(0) = 2(1 - 1/N)$ for a thermal ensemble~\cite{Mandel}] but can instead reach $g_j^{(2)}(0) > 2$.
This is similar to the extreme photon bunching observed in the subradiant regime of previous models~\cite{Meiser3,Shankar,Reilly,Shankar2}, and is explored in more depth in Ref.~\cite{Shankar2} for the traditional superradiant lasing model~\cite{Meiser}.
Although superbunching photon statistics $g^{(2)} (0) > 2$ can arise in unentangled ensembles~\cite{Bojer}, we believe the behavior in Fig.~\ref{LightProperties} is a similar effect to Refs.~\cite{Shankar,Shankar2} where entangled collective dark states form due to the bounded state space of the respective $\mathrm{SU}(2)$ subgroups.
This behavior is consistent with our steady-state inversion results in the inset of Fig.~\ref{LightOutput}(b) where the bounded state space keeps the population near zero even when $W$ and $\Gamma_c$ are different by orders of magnitude. 
Importantly, this signifies a highly non-gaussian steady-state which cannot be accurately modeled using a mean-field approximation~\cite{Temnov}, and instead requires higher-order moments~\cite{Shankar,Shankar2}. 
This exemplifies why our exact quantum master equation simulation technique from Sec.~\ref{Sec:SimTechniques} is so powerful as we have direct access to all-order steady-state moments to accurately describe the light output from nonclassical many-body states.

It should be noted that the $E$ and $J$ observables are symmetric about the potential phase transition at $W = \Gamma_c$.
This follows from a symmetry of the master equation under exchange of the pumping and decay processes.
Specifically, exchanging the labels $\ket{g} \leftrightarrow \ket{e}$ and $\ket{p_x} \leftrightarrow \ket{p_z}$ leaves the master equation, Eq.~\eqref{AtomicMasterEq}, invariant.
Therefore, the strong-pumping regime of the $x$-cavity ($W \gg \Gamma_c$) can be viewed as the weak-pumping regime of the $z$-cavity, and vice versa.
This explains why the superradiant component of the $z$-cavity's intensity shrinks towards zero when $W \gg \Gamma_c$, while the superradiant component of the $x$-cavity shrinks towards zero when $W \ll \Gamma_c$.
We note that single-particle decoherence will, in general, break this symmetry.
For example, spontaneous emission from $\ket{e} \rightarrow \ket{g}$ outside the cavities is not invariant under exchange of the internal-state labels.
We will discuss single-particle decoherence further in Sec.~\ref{Sec:Conclusion}.
In the next two sections, we explore the steady-state interparticle and spin-momentum entanglement generated in our system.

\section{Spin-Momentum Hybrid Entanglement} \label{Sec:SpinMomentumEntanglement}
The superradiant decay process $\hat{\mathcal{D}}[\sqrt{\Gamma_c} \hat{E}_-]$ entangles the spin and $x$-momentum degrees of freedom, as the collective lowering operator induces an $x$-momentum flip with each spin flip.
Similarly, the superradiant pumping process $\hat{\mathcal{D}}[\sqrt{W} \hat{J}_+]$ entangles the spin and $z$-momentum degrees of freedom. Since the spin-$x$-momentum and spin-$z$-momentum hybrid entanglement are symmetric about $W = \Gamma_c$, as for the $E$ and $J$ observables above, we focus here on the spin-$x$-momentum hybrid entanglement.
We note in passing that hybrid entanglement specifically refers to entanglement between discrete and continuous degrees of freedom~\cite{Li}.
We adopt this language here because momentum is physically a continuous variable with an infinite Hilbert space, even if the dynamics in our model are restricted to a discrete set of values.
However, the analysis we perform in this section can also be viewed as a more general class of entanglement, namely, algebraic entanglement between the $J$ and $K$ $\mathfrak{su} (2)$ sub-algebras~\cite{ActuallyJarrodsPaper3,Balachandran,ReyesLega}.

We adopt an entropic approach in this section and define the von Neumann entropy of a density matrix $\hat{\rho}_i$ as
\begin{equation} \label{VN_Entropy}
    S[\hat{\rho}_i] = - \Tr[ \hat{\rho}_i \ln( \hat{\rho}_i) ].
\end{equation}
At first glance, the calculation of the entanglement entropy of the reduced density matrices,
\begin{equation}
    \hat{\rho}_J = \Tr_K [\hat{\rho}], \quad \hat{\rho}_K = \Tr_J [\hat{\rho}],
\end{equation}
where $\Tr_i$ denotes the partial trace over subsystem $i$, appears to be a numerically demanding task. This is because in our many-body system the reduced density matrices typically scale as $(2^N)^2$.
Moreover, states in the $\mathrm{SU}(4)$ bosonic subspace are, in general, not separable with respect to the spin and momentum degrees of freedom~\cite{ActuallyJarrodsPaper3}.

In an accompanying paper~\cite{ActuallyJarrodsPaper3}, we show a deep connection between algebraic entanglement entropy in composite $\mathrm{SU}(2)\otimes\mathrm{SU}(2)$ systems and the irreducible representations of the Lie groups describing the different degrees of freedom.
From this, we develop an algorithm for the partial trace over the spin or momentum subspace that directly maps each reduced density matrix into the polynomial-scaling basis of Sec.~IV of Ref.~\cite{Xu}, which scales as $\mathcal{O}(N^3)$.
Moreover, density matrices in this basis are given by a direct sum of sub-density matrices (i.e., are block diagonal), whose maximum dimensionality is $(N+1)^2$, as each block corresponds to an irreducible representation of $\mathrm{SU}(2)$~\cite{Georgi,Pfeifer}.
In our system, the irreducible representations of the respective $\mathrm{SU}(2)$ subgroups are labeled by the eigenvalues $j(j+1)$ or $k(k+1)$ of the quadratic Casimir operators $\hat{J}^2$ or $\hat{K}^2$ (we switch notation to lowercase $j$ to label the spin dipole length here as to not confuse it with the subsystem label $J$).
Therefore, the reduced density matrices can be written as
\begin{equation}
    \hat{\rho}_J = \bigoplus_{j = 0}^{N / 2} \left[ c_j \bigoplus_{i = 1}^{d_j} \hat{\rho}_j^i \right], \quad \hat{\rho}_K = \bigoplus_{k = 0}^{N / 2} \left[ c_k \bigoplus_{i = 1}^{d_k} \hat{\rho}_k^i \right],
\end{equation}
where the multiplicity of the $\mathrm{SU} (2)$ irreducible representations is given by~\cite{Mandel}
\begin{equation} \label{d_l}
    d_l = \frac{N! (2 l +1)}{(\frac{N}{2} + l + 1)! \ (\frac{N}{2} - l)!}, \quad l \in \{ j,k \}.
\end{equation}
Since the reduced density matrices are block diagonal, their eigenvalues can be obtained in a block-by-block manner to compute the entanglement entropies. 
Namely, one only needs to diagonalize the sub-density matrices of a single copy of each $\mathrm{SU}(2)$ irreducible representation (i.e., $\hat{\rho}_j^1$ and $\hat{\rho}_k^1$), which greatly reduces the numerical complexity of the calculation.
We present pseudocode for the algebraic entanglement entropy algorithm in Appendix~\ref{Appendix:Algorithm} (see Algorithm~\ref{alg:EE}), and refer the reader to Ref.~\cite{ActuallyJarrodsPaper3} for further details.

While a nonzero entanglement entropy of the reduced density matrices, $S[\hat{\rho}_J]$ and $S[\hat{\rho}_K]$, is sufficient to demonstrate hybrid entanglement for pure states, this is no longer true for mixed states due to entanglement between the system and environment~\cite{Horodecki}.
As an alternative, one can instead calculate the conditional quantum entropy~\cite{Cerf}.
\begin{equation}
    S(X \lvert Y)[\hat{\rho}] = S[\hat{\rho}] - S[\hat{\rho}_Y], \quad X \neq Y \in \{ J,K \}.
\end{equation}
Classically, the conditional entropy in positive-definite~\cite{Cerf}, and so an entropic witness of entanglement between subsystems is given by
\begin{equation} \label{CoherentInformation}
    S(X \lvert Y)[\hat{\rho}] < 0, \quad X \neq Y \in \{ J,K \}.
\end{equation}
This is a necessary and sufficient condition for entanglement in pure states, but only a sufficient condition for mixed states, as studied here.
Therefore, $S(J \lvert K)$ may be negative while $S(K \lvert J)$ is positive for an entangled state, and vice versa, or both may be positive for an entangled state.
The negativity of the conditional entropy is known as coherent information~\cite{Wilde}.
\begin{equation}
    I(X \rangle Y)[\hat{\rho}] = - S(X \lvert Y)[\hat{\rho}] = S[\hat{\rho}_Y] - S[\hat{\rho}],
\end{equation}
and is directly connected to the ability of a quantum state to serve as a resource in entanglement-enhanced information processing, such as quantum dense coding~\cite{Prabhu}.
One can think of coherent information as playing a role for quantum correlations analogous to that of mutual information in the classical case~\cite{Wilde}.
Direct witnesses for coherent information were proposed in Ref.~\cite{Vempati}.

Since calculating the von Neumann entropy requires the full spectrum of the density matrices, it is no longer sufficient to simulate only the diagonal blocks $\hat{\rho}_{\ell}$ of Eq.~\eqref{rhoDecomp}; one must also track the coherences between blocks, $\hat{\rho}_c$, to obtain the correct eigenvalues and eigenvectors.
Therefore, we now perform our simulations in the traditional Schwinger-boson basis presented in Ref.~\cite{ActuallyJarrodsPaper}.
Because calculations can no longer be done block by block in this basis, the Liouville space scales as $\mathcal{O}(N^6)$, which limits the total number of atoms we can simulate.

\begin{figure}
    \centerline{\includegraphics[width=0.75\linewidth]{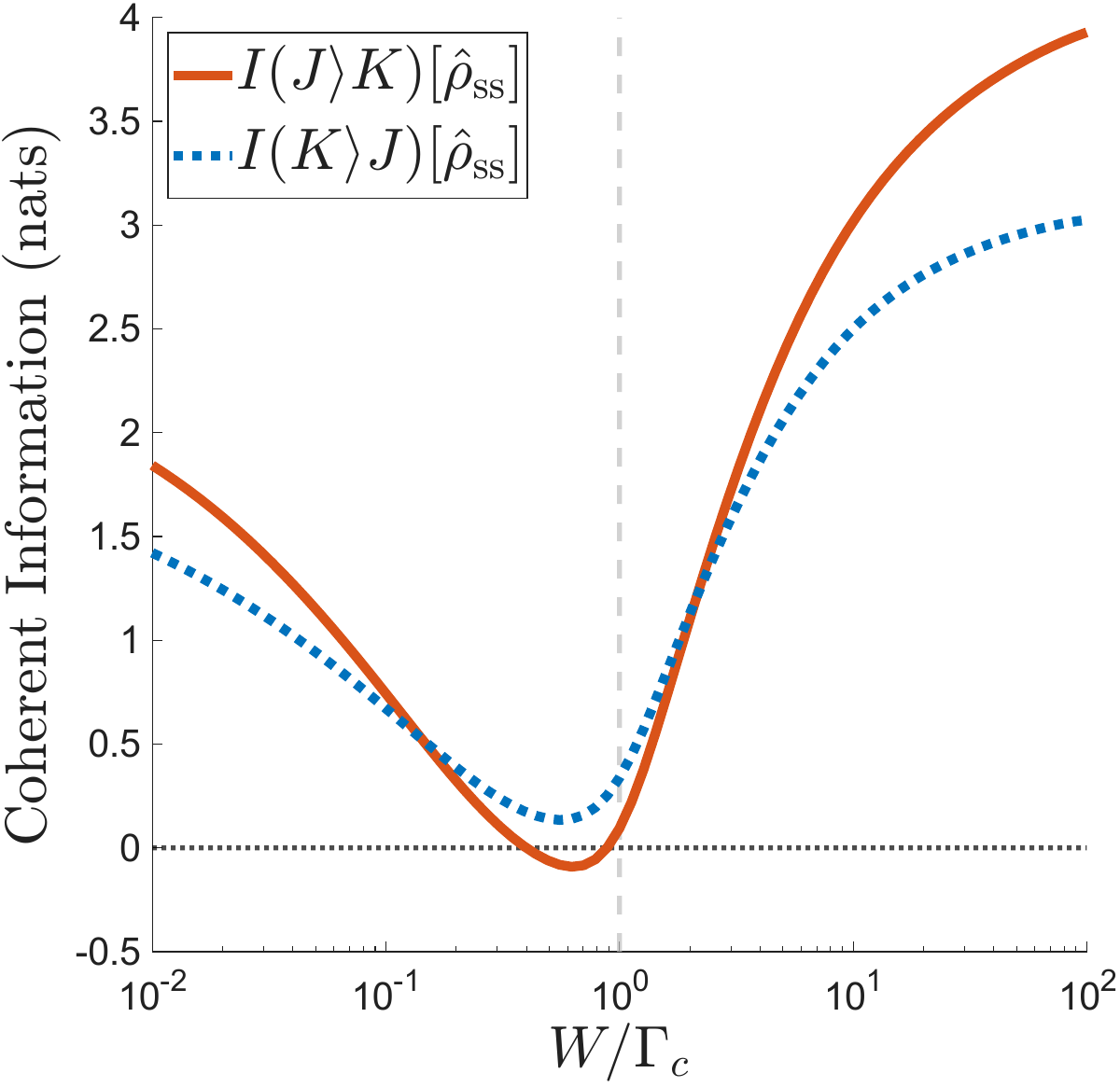}}
    \caption{Steady-state coherent information $I(X \rangle Y)[\hat{\rho}_{\mathrm{ss}}]$ from Eq.~\eqref{CoherentInformation} for the subspaces $X \neq Y \in \{ J, K \}$. We use the algorithm from Ref.~\cite{ActuallyJarrodsPaper3} with $N = 10$. The dotted black line marks $I(X \rangle Y)[\hat{\rho}_{\mathrm{ss}}] = 0$; any point above it indicates quantum correlations (hybrid entanglement) between the spin and $x$-momentum degrees of freedom.}
    \label{CohInfoFig}
\end{figure}
In Fig.~\ref{CohInfoFig}, we show the steady-state coherent informations $I(J \rangle K)[\hat{\rho}_{\mathrm{ss}}]$ and $I(K \rangle J)[\hat{\rho}_{\mathrm{ss}}]$ for $N = 10$.
The dotted black line marks $I(X \rangle Y)[\hat{\rho}_{\mathrm{ss}}] = 0$, so any point above it indicates quantum correlations (hybrid entanglement) between spin and momentum.
We see that when $W \ll \Gamma_c$ or $W \gg \Gamma_c$, a significant amount of coherent information between the two subsystems exists at steady state.
For example, at $W \approx 10^2 \, \Gamma_c$, $I(J \rangle K)[\hat{\rho}_{\mathrm{ss}}] \sim 4~\mathrm{nats}$ and $I(K \rangle J)[\hat{\rho}_{\mathrm{ss}}] \sim 3~\mathrm{nats}$ which are both roughly half the value of a maximally entangled state, $N \ln 2 \approx 6.93$.
Interestingly, the coherent information continues to vary away from $W = \Gamma_c$ even when the spin inversion (and the intensity and dipole length in one regime) remain relatively constant.
Since the spin-$z$-momentum coherent information is reflected about $W = \Gamma_c$, Fig.~\ref{CohInfoFig} also shows that the spin maintains significant quantum correlations with multiple external degrees of freedom away from the critical point, which may be a useful resource for quantum information applications.
When $W \approx \Gamma_c$, the coherent informations become small, with $I(J \rangle K)[\hat{\rho}_{\mathrm{ss}}]$ even dropping below zero.
In Ref.~\cite{ActuallyJarrodsPaper3}, we show that the total entropy of the steady state, $S[\hat{\rho}_{\mathrm{ss}}]$, increases with $W$ for $W < \Gamma_c$ and decreases for $W > \Gamma_c$, reaching a maximum at $W = \Gamma_c$ (about which $S[\hat{\rho}_{\mathrm{ss}}]$ is symmetric).
Thus, it is likely that this maximal entanglement with the environment suppresses spin-momentum quantum correlations near the point $W = \Gamma_c$.

It may be surprising that the coherent information is larger for $W \gg \Gamma_c$ than for $W \ll \Gamma_c$, since $\hat{E}_-$ directly entangles spin and $x$-momentum, while $\hat{J}_+$ does not.
One possible explanation is that most, or all, trajectories (in the Monte-Carlo wave function picture; see Sec.~\ref{Sec:Metrology}) have larger entanglement entropy when $W \ll \Gamma_c$ than when $W \gg \Gamma_c$, but the spin-$x$–momentum configuration is more “consistent” per trajectory for $W \gg \Gamma_c$, as it is a perturbation about a degenerate eigenspace of~$\hat{J}_+$.
Consequently, more spin-$x$-momentum quantum correlations survive when averaging over trajectories in the regime $W \gg \Gamma_c$.
Further investigation of this behavior is left for future work.
Moreover, future studies will examine the scaling of spin-momentum quantum correlations with $N$, which is currently limited by the need to diagonalize the full density matrix $\hat{\rho}_{\mathrm{ss}}$.
We now turn from quantum information science to quantum metrology and assess the system's usefulness for quantum-enhanced sensing.

\section{Interparticle Entanglement} \label{Sec:Metrology}
We now assume that the light emitted from the two cavities is monitored by separate photodetectors.
In this case, the system no longer evolves as a mixed state, since measurement of the output fields projects out the entangled superpositions between different system-environment configurations~\cite{Holland}.
The system then follows a specific quantum trajectory, which can be simulated using the Monte-Carlo wave function method~\cite{Molmer,Meystre}.
Here, the state vector $\ket{\psi}$ is independently evolved under the non-Hermitian Schr\"odinger equation
\begin{equation} \label{NHSchrodingerEq}
    \pd{} \ket{\psi} = - \frac{i}{\hbar} \hat{H}_{\mathrm{NH}} \ket{\psi},
\end{equation}
with the non-Hermitian Hamiltonian 
\begin{equation} \label{H_NH}
    \hat{H}_{\mathrm{NH}} = - \frac{i \hbar}{2} \left( W \hat{J}_- \hat{J}_+ + \Gamma_c \hat{E}_+ \hat{E}_- \right).
\end{equation}
Since the dynamics under Eq.~\eqref{NHSchrodingerEq} are nonunitary, the norm of the state vector decays over time.
Quantum jumps can then be stochastically simulated by comparing the norm to a random number $r \in (0,1)$; a jump occurs when $\ip{\psi}{\psi} \leq r$.
The type of jump is also chosen stochastically, weighted by $\delta p_i / \delta p$, where $\delta p_i \propto \bra{\psi} \hat{O}_i^{\dagger} \hat{O}_i \ket{\psi}$ for the jump operators $\hat{O}_i \in \{ \sqrt{W} \hat{J}_+, \sqrt{\Gamma_c} \hat{E}_- \}$, and $\delta p = \sum_i \delta p_i$~\cite{Meystre}.
The Monte-Carlo wave function method reproduces Eq.~\eqref{AtomicMasterEq} when averaged over many trajectories~\cite{Molmer}, but here it represents a single realization of the experiment while monitoring light emission from the two cavities.

As a witness of interparticle entanglement, we calculate the quantum Fisher information (QFI) with respect to different operators.
The QFI quantifies the maximal metrological usefulness of a quantum state for a parameter encoded by the operator, and a QFI exceeding $N$ provides a sufficient condition for interparticle entanglement in $\mathrm{SU}(n)$ systems~\cite{Reilly2}.
Rather than computing the QFI for a single operator, we can determine the maximal achievable QFI for a given state by evaluating the eigenvalues of the QFI matrix (QFIM)~\cite{Reilly2}.
\begin{equation} \label{QFIMeigvals}
    \bm{\mathcal{F}} \vec{\mathcal{O}}_i = \lambda_{i} \vec{\mathcal{O}}_i.
\end{equation}
Here, the elements of the QFIM are given by~\cite{Liu2} 
\begin{equation} \label{QFIM}
    \bm{\mathcal{F}}_{\mu \nu} = 2 \exv{\hat{G}_{\mu} \hat{G}_{\nu} + \hat{G}_{\nu} \hat{G}_{\mu}} - 4 \exv{\hat{G}_{\mu}} \exv{\hat{G}_{\nu}},
\end{equation}
for pure states, where $\hat{G}_{\mu}$ is an operator in an orthonormal basis [see Eq.~\eqref{OpBasis}].
Finding the parameter $\Phi$ to which a prepared state is most sensitive is then equivalent to finding the maximal eigenvalue $\lambda_{\mathrm{max}}$ of the QFIM~\cite{Reilly2}, which gives the maximal attainable QFI for the state.
The eigenvector $\vec{\mathcal{O}}_{\mathrm{max}}$ corresponding to $\lambda_{\mathrm{max}}$ is directly related to the optimal generator $\hat{\mathcal{G}}$ that encodes the parameter, $\exp[-i \hat{\mathcal{G}} \Phi]$, via a sum over the coefficients of the orthonormal operator basis~\cite{Reilly2},
\begin{equation} \label{OptGenerator}
    \hat{\mathcal{G}} = \mathcal{O}_{\mathrm{max}}^{\mu} \hat{G}_{\mu}. 
\end{equation}
Here, we have adopted Einstein's summation notation and the vector notation $\vec{\mathcal{O}_i} = \mathcal{O}_i^{\mu}$. 

The $\mathrm{SU}(4)$ group has six $\mathfrak{su}(2)$ sub-algebras with associated raising operators 
\begin{equation}
    \begin{aligned}
& \hat{\mathcal{Q}}^+ = \sum_j \op{e,r}{g,l}_j, \quad & \hat{\Sigma}^+ = \sum_j \op{e,l}{g,r}_j, \\ 
&\hat{\mathcal{M}}^+ = \sum_j \op{e,r}{g,r}_j, \quad & \hat{\mathcal{N}}^+ = \sum_j \op{e,l}{g,l}_j, \\ 
& \hat{\mathcal{U}}^+ = \sum_j \op{e,r}{e,l}_j, \quad & \hat{\mathcal{V}}^+ = \sum_j \op{g,r}{g,l}_j. 
    \end{aligned}
\end{equation}
Each raising operator gives the Hermitian components of each algebra $\hat{O}_x$, $\hat{O}_y$, and $\hat{O}_z$ according to Eq.~\eqref{SU2Generators}.
We can then use these Hermitian operators to create a 15-dimensional orthonormal operator basis~\cite{Reilly2},
\begin{equation} \label{OpBasis}
    \begin{aligned}
\hat{G}_{\mu} \in \{ & \hat{\mathcal{Q}}_x, \hat{\mathcal{Q}}_y, \hat{\mathcal{Q}}_z, \hat{\Sigma}_x, \hat{\Sigma}_y, \hat{\Sigma}_z, \hat{\mathcal{M}}_x, \hat{\mathcal{M}}_y, \\
& \hat{\mathcal{N}}_x, \hat{\mathcal{N}}_y, \hat{\mathcal{P}}_z, \hat{\mathcal{U}}_x, \hat{\mathcal{U}}_y, \hat{\mathcal{V}}_x, \hat{\mathcal{V}}_y \},
    \end{aligned}
\end{equation}
where $\hat{\mathcal{P}}_z = ( \hat{\mathcal{M}}_z - \hat{\mathcal{N}}_z ) / \sqrt{2}$.
Using this operator basis, we can now compute the QFIM throughout the state vector evolution using Eq.~\eqref{QFIM} in order to find the maximum QFI $\lambda_{\mathrm{max}}$. 
Since the state is, in general, not normalized under the evolution of the non-Hermitian Hamiltonian, we note that one has to use the normalized state $\ket{\psi' (t)} = \ket{\psi (t)} / \sqrt{\ip{\psi (t)}{\psi (t)}}$ to compute the QFIM at each time step. 

\begin{figure}
    \centerline{\includegraphics[width=\linewidth]{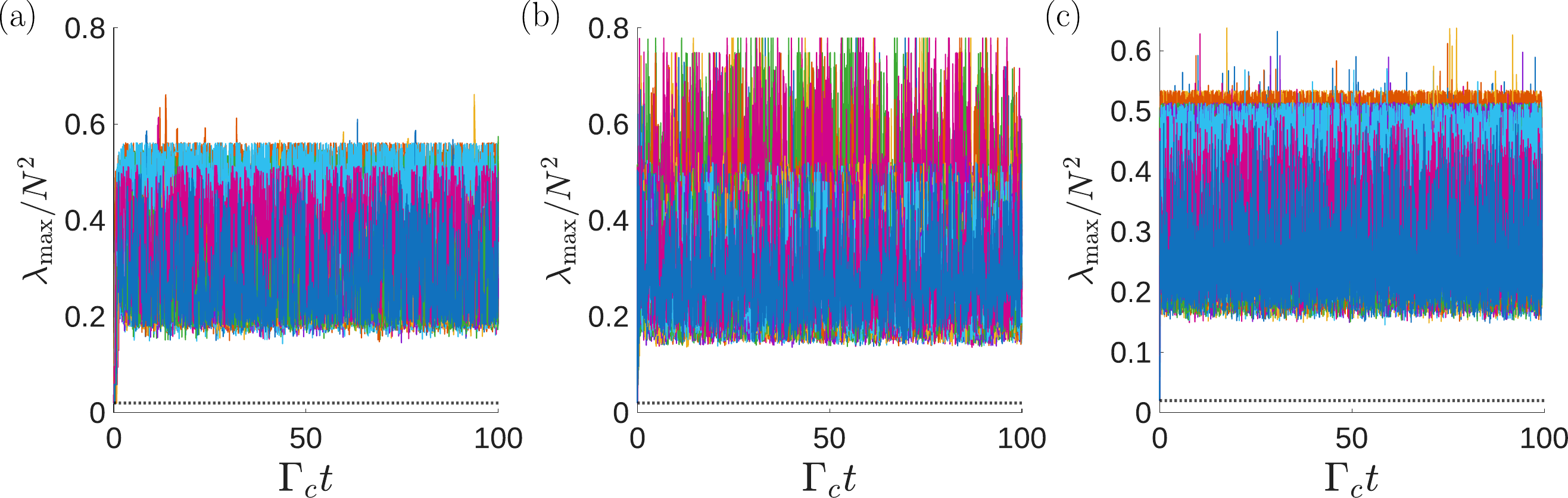}}
    \caption{The maximum quantum Fisher information $\lambda_{\mathrm{max}}$ for $N_{\mathrm{traj}} = 50$ different Monte-Carlo wave function trajectories.
    The different trajectories are signified by different colors, while the dotted gray line indicates the standard quantum limit $\lambda_{\mathrm{max}} = N$. 
    Each figure has $N = 50$, and the pump values are given by (a) $W = 0.1 \Gamma_c$, (b) $W = \Gamma_c$, and (c) $W = 10 \Gamma_c$.}
    \label{QuTrajFig}
\end{figure}
In Fig.~\ref{QuTrajFig}, we show the maximum QFI, $\lambda_{\mathrm{max}}$, for (a) $W = 0.1 \Gamma_c$, (b) $W = \Gamma_c$, and (c) $W = 10 \Gamma_c$ with $N = 50$ throughout the evolution.
Each plot shows $N_{\mathrm{traj}} = 50$ trajectories to demonstrate that interparticle entanglement is generated in every run.
The dotted gray lines indicate the standard quantum limit (SQL), $\lambda_{\mathrm{max}} = N$, which sets the fundamental sensitivity threshold from shot noise in uncorrelated quantum states.
Any QFI exceeding the SQL is thus a sufficient witness of interparticle entanglement.
We see that, after the initial superradiant burst (i.e., after $\hat{\rho}$ would reach steady-state), the system in every trajectory consistently remains interparticle-entangled with $\lambda_{\mathrm{max}} > 0.15 N^2$.
This implies that the mixed state
\begin{equation}
    \hat{\rho} (t) = \frac{1}{N_{\mathrm{traj}}} \sum_{j = 1}^{N_{\mathrm{traj}}} \op{\psi'_j (t)}{\psi'_j (t)}
\end{equation}
also contains interparticle entanglement, although its $\lambda_{\mathrm{max}}$ may not exceed the SQL because different trajectories have different optimal generators $\hat{\mathcal{G}}$ at a given time.

\begin{figure}
    \centerline{\includegraphics[width=\linewidth]{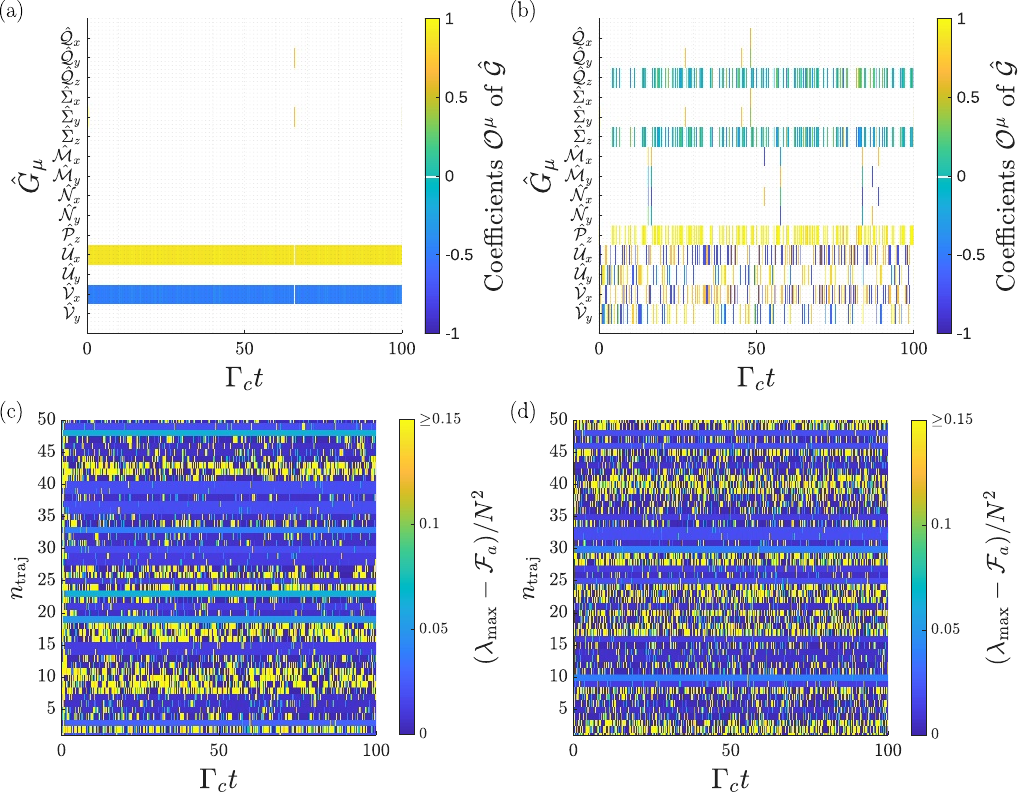}}
    \caption{(a) and (b) show the coefficients $\mathcal{O}^{\mu}$ of the optimal generator $\hat{\mathcal{G}}$ [Eq.~\eqref{OptGenerator}] for two sample trajectories ($n_{\mathrm{traj}} = 10$ and $40$) with $N = 50$ and $W = 10 \Gamma_c$. The orthonormal operator basis $\hat{G}_{\mu}$ is given in Eq.~\eqref{OpBasis}. (c) and (d) show the difference between the maximum quantum Fisher information $\lambda_{\mathrm{max}}$ and the quantum Fisher information for acceleration, $\mathcal{F}_a$, for $N_{\mathrm{traj}} = 50$ Monte-Carlo trajectories. The parameters are $N = 50$ with (c) $W = 0.1 \Gamma_c$ and (d) $W = 10 \Gamma_c$.}
    \label{KzFig}
\end{figure}
To this point, we can examine the optimal generator $\hat{\mathcal{G}}$ across different trajectories.
In Figs.~\ref{KzFig}(a) and~(b), we show the coefficients of the optimal generator for two trajectories with $N = 50$ and $W = 10 \Gamma_c$ (trajectory numbers $n_{\mathrm{traj}} = 10$ and $40$, respectively).
These trajectories illustrate the general behavior we observe in all runs.
In particular, some trajectories are dominated by $\hat{\mathcal{G}} \sim \hat{\mathcal{U}}_x - \hat{\mathcal{V}}_x$, as in Fig.~\ref{KzFig}(a).
Other trajectories do not maintain a consistent $\hat{\mathcal{G}}$, with the optimal generator switching between linear combinations of ${ \hat{\mathcal{Q}}_z, \hat{\Sigma}_z, \hat{\mathcal{P}}_z, \hat{\mathcal{U}}_x, \hat{\mathcal{U}}_y, \hat{\mathcal{V}}_x, \hat{\mathcal{V}}_y }$, as in Fig.~\ref{KzFig}(b).
To gauge how many trajectories exhibit the behavior in Fig.~\ref{KzFig}(a) versus Fig.~\ref{KzFig}(b), we note that $[\hat{\mathcal{U}}_i, \hat{\mathcal{V}}_j] = 0$ for $i,j \in \{ x,y,z \}$.
Therefore, performing the rotation
\begin{equation} \label{RotToKz}
    \ket{\tilde{\psi} (t)} = \hat{U}_a \ket{\psi' (t)} = \exp[- i (\hat{\mathcal{U}}_y - \hat{\mathcal{V}}_y) \pi / 2] \ket{\psi' (t)},
\end{equation}
rotates the optimal generator from $\hat{\mathcal{G}} \sim \hat{\mathcal{U}}_x - \hat{\mathcal{V}}_x$ to $\tilde{\hat{\mathcal{G}}} \sim \hat{\mathcal{U}}_z + \hat{\mathcal{V}}_z = \hat{K}_z$. 
Sensitivity to the $\hat{K}_z$ operator can be seen as a sensitivity to an acceleration $a$ along the $x$-direction through
\begin{equation} \label{PhaseEncoding}
    \frac{\hat{p}^2}{2 m} - m a \hat{x} \longrightarrow \frac{\hat{p}^2}{2 m} + a \tau \hat{p} \Longrightarrow a \tau \hat{K}_z,
\end{equation}
where $\rightarrow$ denotes moving into the co-accelerating reference frame~\cite{Reilly3,ActuallyJarrodsPaper2}, $\Rightarrow$ denotes mapping onto the $\mathrm{SU}(4)$ basis via Appendix~\ref{Appendix:MomentumReduction}, and identity terms have been dropped in each expression.
The encoding time $\tau$ corresponds to the interval during which the atoms are in free flight under acceleration after the atom-cavity interactions are turned off (e.g., atoms exiting the cavities in an atomic beam setup~\cite{Liu,Fama}), and it is assumed to be well known.
The QFI for acceleration is then given by the scaled variance,
\begin{equation}
    \mathcal{F}_a = 4 \left[ \Delta \left( \hat{K}_z / \sqrt{2} \right) \right]^2,
\end{equation}
where the $\sqrt{2}$ factor normalizes $\hat{K}_z$ in the orthonormal operator basis of Eq.~\eqref{OpBasis}, putting it on the same footing as the results of Fig.~\ref{QuTrajFig}~\cite{Reilly2}.
In Figs.~\ref{KzFig}(c) and~(d), we show the difference between the maximum QFI and the QFI for acceleration, $\lambda_{\mathrm{max}} - \mathcal{F}_a$, for each trajectory after rotating the state via Eq.~\eqref{RotToKz}.
Figure~\ref{KzFig}(c) corresponds to $W = 0.1 \Gamma_c$ and Fig.~\ref{KzFig}(d) to $W = 10 \Gamma_c$. In both cases, many trajectories exhibit the behavior of Fig.~\ref{KzFig}(a), corresponding to runs that are predominantly blue or green in Figs.~\ref{KzFig}(c) and~(d).
Meanwhile, runs that are predominantly yellow correspond to trajectories exhibiting the behavior of Fig.~\ref{KzFig}(b).
Steady-state points that are not bright yellow must have QFI beyond the SQL, since Fig.~\ref{QuTrajFig} always shows $\lambda_{\mathrm{max}} \geq 0.15 N^2$, indicating points where quantum-enhanced accelerometry can potentially be performed.
As discussed in Appendix~\ref{Appendix:Pyramid}, the state that maximizes $\lambda_{\mathrm{max}}$ for a trajectory exhibiting the behavior of Fig.~\ref{KzFig}(a) is rotated under $\hat{U}_a$ to a state with NOON-like, or more generally GHZ-like (Greenberger–Horne–Zeilinger), properties~\cite{Degen,Pezze} in the $K$ degree of freedom [see Figs.~\ref{PyramidOverlap} and~\ref{LayerPops}].
This helps explain its extreme sensitivity to $\hat{K}_z$ in these runs, $\mathcal{F}_a \sim N^2 / 2$, and exemplifies the power of the optimal generator protocol from Ref.~\cite{Reilly2} at finding metrologically useful entangled states.

Interestingly, while both $W = 0.1 \Gamma_c$ and $W = 10 \Gamma_c$ have many trajectories behaving like Fig.~\ref{KzFig}(a), we do not observe this behavior at $W = \Gamma_c$.
This suggests that the behavior of Fig.~\ref{KzFig}(a) may result from trajectories becoming trapped in a dark subspace, similar to the discussion in Sec.~\ref{Sec:Superradiance}, where different states in the subspace remain sensitive to the same operator (albeit with different sensitivities, as shown in Fig.~\ref{QuTrajFig}). Moreover, as discussed in Sec.~\ref{Sec:SpinMomentumEntanglement}, spin-momentum hybrid entanglement is minimized near $W \sim \Gamma_c$ while entanglement with the environment is maximized, indicating that spin-momentum correlations may play a key role in the system's sensitivity to acceleration.
We leave further investigation of this feature to future work.

\section{Conclusion and Outlook} \label{Sec:Conclusion}
In this paper, we introduced a cross-bad-cavity system whose interaction with a cloud of atoms leads to fully collective steady-state superradiance.
Unlike superradiance in $\mathrm{SU} (2)$ systems, the resulting multi-level superradiant states in this system show clear signatures of nonclassical properties [we contrast our full master equation simulation results to results from a mean-field simulation in Appendix~\ref{Appendix:MF}], such as super-Poissonian photon statistics and steady-state entanglement. 
We demonstrated that $W = \Gamma_c$ marks the critical point of a potential dissipative phase transition where the intensity and intensity fluctuations of the light output from the $x$-cavity and $z$-cavity, as well as the inversion and quadratic Casimir operators, switch behavior.
We also found that the superradiant bursts and associated momentum kicks generate both interparticle and spin-momentum entanglement, which can potentially be harnessed for quantum-enhanced sensing and quantum information processing. Moreover, the system provides a rich platform for many-body physics, as a subspace of dark states emerges, leading to nontrivial interference that can potentially be exploited for quantum-enhanced accelerometry after postselection.

Future work can focus on potentially ``steering'' the system toward these dark states to increase the fraction of trajectories with significant spin-momentum entanglement capable of quantum-enhanced sensing.
It would also be interesting to explore connections between our system and spin-momentum entanglement generated via superradiance in the experiments of Refs.~\cite{Finger,Chelpanova}, as well as the cross-cavity setups of Refs.~\cite{Leonard,Leonard2,Lang,Baumgartner}.

Complementary work could investigate the potential of the generated entanglement as a resource for quantum technologies. In particular, the hybrid spin-momentum entanglement discussed in Sec.~\ref{Sec:SpinMomentumEntanglement} may prove useful for both quantum metrology and quantum information processing. In the context of metrology, it could enable ancilla-assisted sensing, where a phase encoded in one degree of freedom is inferred by measuring the other~\cite{ActuallyJarrodsPaper}, potentially avoiding destructive measurements and improving experimental efficiency. Beyond metrology, hybrid spin-momentum entanglement may also serve as a valuable resource for quantum information processing, with potential applications including entanglement swapping~\cite{Andersen,Takeda,Tang,Adhikari,Yang,Bayal}, quantum networking~\cite{Chang,Sheng}, quantum teleportation~\cite{Park,Lee}, and quantum key distribution~\cite{Bose,Fujiwara}.

A rigorous assessment of the potential utility of the generated entanglement should also include a detailed investigation of the detrimental effects of single-particle decoherence.
We note that the master equation at the center of our analysis, Eq.~\eqref{AtomicMasterEq}, does not include single-particle emission or dephasing.
This is typically valid in the superradiant regime for cavities with large collective cooperativity parameters, $N \mathcal{C}_x = N \Gamma_c / \gamma_e \gg 1$ and $N \mathcal{C}_z = N g_z^2 / (\kappa_z \gamma_a) \gg 1$, where $\gamma_e$ and $\gamma_a$ are the natural linewidths of the states $\ket{e}$ and $\ket{a}$, respectively. This assumption is crucial for experimental implementation, as spontaneous photon emission can remove the atomic cloud from the bosonic $\mathrm{SU}(4)$ subspace (reducing its quartic Casimir eigenvalue) and also from the discrete momentum grid, potentially lowering spin-momentum entanglement.
Notably, achieving large single-atom cooperativities $\mathcal{C} \gg 1$ in cavity systems has attracted significant experimental and theoretical interest recently~\cite{Kawasaki,Ren,Kroeze2,Bin,Codreanu}, making implementation of our system without adverse effects from spontaneous emission feasible.
Further analysis of single-particle decoherence on the metrologically useful collective dark states is left for future work.

Lastly, as noted in Sec.~\ref{Sec:Metrology}, it would be interesting to consider our system in an atomic beam setup, similar to Refs.~\cite{Liu,Jager2,Jager3,Fama}, but now through two cavities and without initial excitation.
In this picture, the system acts as a continuous entanglement generator: atoms inside the cavities become entangled with one another, generating both interparticle and hybrid spin-momentum entanglement, while new atoms enter to replace those exiting.
This naturally implements the ``turning off'' of the atom-cavity interactions, allowing atoms in free space to evolve solely under the phase-encoding Hamiltonian of Eq.~\eqref{PhaseEncoding} for quantum-enhanced accelerometry.
With new atoms continuously replacing measured ones, such a setup has the potential to enable continuous readout of acceleration.

\section*{Acknowledgments}
We would like to thank John Cooper and Charles Marrder for useful discussions. 
This material is based upon work supported by the U.S. Department of Energy, Office of Science, National Quantum Information Science Research Centers, Quantum Systems Accelerator (Award No. DE-SCL0000121), and by the National Science Foundation Grant Nos.\ 2016244 and 2317149.
G.W.H acknowledges funding by the German Ministry of Research, Technology and Space (BMFTR, Project ``NiQ: Noise
in Quantum Algorithms'') and by the Deutsche Forschungsgemeinschaft (DFG, German Research Foundation) under project number 525057097 (``Quantum Many-Body Dynamics of Matter and Light in Cavity-QED'').
S.B.J. acknowledges support from the Deutsche Forschungsgemeinschaft (DFG, German Research Foundation) under Projects DFG No. 277625399 -- TRR 185 OSCAR (``Open System Control of Atomic and Photonic Matter'', B4), and DFG No. 277146847 -- CRC 1238 (``Control and dynamics of quantum materials", C05) and under Germany’s Excellence Strategy -- Cluster of Excellence Matter and Light for Quantum Computing (ML4Q), Grant No. EXC 2004/1 -- 390534769.

\appendix
\renewcommand{\thefigure}{\thesection\arabic{figure}}
\counterwithin{figure}{section}

\begin{widetext}
\section{Derivation of the Model} \label{Appendix:ModelDerivation}
In this Appendix, we derive the model presented in Eq.~\eqref{AtomicMasterEq}. 

\subsection{Original master equation}
We begin with the Hamiltonian
\begin{equation} \label{H_0}
    \begin{aligned}
\hat{H}_0 = & \hbar \omega_x \hat{a}_x^{\dagger} \hat{a}_x + \hbar \omega_z \hat{a}_z^{\dagger} \hat{a}_z + \sum_{j = 1}^N \Bigg[ \frac{\vec{p}_j \cdot \vec{p}_j}{2 m} + \hbar \omega_e \op{e}{e}_j + \hbar \omega_a \op{a}{a}_j + \frac{\hbar \Omega}{2} \left( \op{g}{a}_j e^{i \omega_l t} + \mathrm{H.c.} \right) \\
&+ \hbar g_x \cos \left( k_x \hat{x}_j \right) \left( \op{e}{g}_j \hat{a}_x + \mathrm{H.c.} \right) + \hbar g_z \cos \left( k_z \hat{z}_j \right) \left( \op{a}{e}_j \hat{a}_z + \mathrm{H.c.} \right) \Bigg],
    \end{aligned}
\end{equation}
where $\omega_x$ and $\omega_z$ are the frequencies of the respective cavities with annihilation operators $\hat{a}_x$ and $\hat{a}_z$, $\omega_e$ and $\omega_a$ are the respective transition frequencies of the $\ket{e}$ and $\ket{a}$ states with respect to $\ket{g}$, and $\omega_l$ is the frequency of the coherent drive $\Omega$. 
Moreover, for atom $j$ with mass $m$, $\hat{x}_j$ and $\hat{z}_j$ are position operators while $\vec{p}_j = (\hat{p}_{x,j}, \hat{p}_{y,j}, \hat{p}_{z,j})^T$ is the total momentum operator.
We also consider dissipative effects that come from the cavity modes decaying into free space at rates $\kappa_x$ and $\kappa_z$. 
This leads to a master equation of the system's density matrix $\hat{\rho}_0$ under the Born-Markov approximation~\cite{Steck},
\begin{equation} \label{rho0_ME}
    \pd{\hat{\rho}_0} = - \frac{i}{\hbar} \left[ \hat{H}_0, \hat{\rho}_0 \right] + \hat{\mathcal{D}} \left[ \sqrt{\kappa_x} \hat{a}_x \right] \hat{\rho}_0 + \hat{\mathcal{D}} \left[ \sqrt{\kappa_z} \hat{a}_z \right] \hat{\rho}_0,
\end{equation}
with the Lindblad superoperator,
\begin{equation} \label{LindbladSuperOp}
    \hat{\mathcal{D}} [ \hat{O} ] \hat{\rho} = \hat{O} \hat{\rho} \hat{O}^{\dagger} - \frac{1}{2} ( \hat{O}^{\dagger} \hat{O} \hat{\rho} + \hat{\rho} \hat{O}^{\dagger} \hat{O} ).
\end{equation}

We now move into an interaction picture that induces the rotation $\hat{\rho}_0 \rightarrow \hat{\rho}_1 = \exp[ i \hat{H}' t/\hbar] \hat{\rho}_0 \exp[ -i \hat{H}' t/\hbar]$ on the density matrix with  the Hamiltonian $\hat{H}' = \sum_j \hbar \omega_l \op{a}{a}_j + \hbar \omega_e \left( \op{e}{e}_j + \hat{a}_x^{\dagger} \hat{a}_x \right) + \hbar \left( \omega_l - \omega_e \right) \hat{a}_z^{\dagger} \hat{a}_z$.
We find 
\begin{equation}
    \begin{aligned}
\hat{H}_1 = & - \hbar \Delta_x \hat{a}_x^{\dagger} \hat{a}_x - \hbar \Delta_z \hat{a}_z^{\dagger} \hat{a}_z + \sum_{j = 1}^N \Bigg[ \frac{\vec{p}_j \cdot \vec{p}_j}{2 m} - \hbar \Delta_a \op{a}{a}_j + \frac{\hbar \Omega}{2} \left( \op{g}{a}_j + \mathrm{H.c.} \right) \\
& + \hbar g_x \cos \left( k_x \hat{x}_j \right) \left( \op{e}{g}_j \hat{a}_x + \mathrm{H.c.} \right) + \hbar g_z \cos \left( k_z \hat{z}_j \right) \left( \op{a}{e}_j \hat{a}_z + \mathrm{H.c.} \right) \Bigg],
    \end{aligned}
\end{equation}
where we have defined the detunings $\Delta_x = \omega_e - \omega_x$, $\Delta_z = \omega_l - \omega_e - \omega_z$, and $\Delta_a = \omega_l - \omega_a$.
The Lindblad superoperator Eq.~\eqref{LindbladSuperOp} is invariant to a rotation of the jump operator, and so the master equation is simply Eq.~\eqref{rho0_ME} with $\hat{\rho}_0 \rightarrow \hat{\rho}_1$ and $\hat{H}_0 \rightarrow \hat{H}_1$. 

\subsection{Elimination of the auxiliary state}
We now assume the auxiliary state $\ket{a}$ is weakly coupled, $\abs{\Delta_a} \gg \sqrt{N} g_z, \abs{\Omega}$, such that it may be adiabatically eliminated over a coarse-grained timescale.
This gives the effective Hamiltonian~\cite{Reiter}
\begin{equation} \label{H_tilde}
    \begin{aligned}
\tilde{\hat{H}} = & - \hbar \Delta_x \hat{a}_x^{\dagger} \hat{a}_x - \hbar \Delta_z \hat{a}_z^{\dagger} \hat{a}_z + \sum_{j = 1}^N \left[ \frac{\vec{p}_j \cdot \vec{p}_j}{2 m} + \hbar g_x \cos \left( k_x \hat{x}_j \right) \left( \hat{\sigma}_j^+ \hat{a}_x + \mathrm{H.c.} \right) + \frac{\hbar g_z \Omega}{2 \Delta_a} \cos \left( k_z \hat{z}_j \right) \left( \hat{\sigma}_j^- \hat{a}_z + \mathrm{H.c.} \right) \right] \\
& + \sum_{j = 1}^N \left[ \frac{\hbar \Omega^2}{4 \Delta_a} \op{g}{g}_j + \frac{\hbar g_z^2}{\Delta_a} \cos^2 \left( k_z \hat{z}_j \right) \op{e}{e}_j \hat{a}_z^{\dagger} \hat{a}_z \right],
    \end{aligned}
\end{equation}
with the Pauli operator $\hat{\sigma}_j^+ = \op{e}{g}_j = \left( \hat{\sigma}_j^-\right)^{\dagger}$.
The last line represents AC Stark shifts which we now ignore based on a similar argument to Ref.~\cite{Reilly} when these shifts are small compared to $\kappa_x$ and $\kappa_z$.
We then find
\begin{equation}
    \begin{aligned}
\tilde{\hat{H}} = - \hbar \Delta_x \hat{a}_x^{\dagger} \hat{a}_x - \hbar \Delta_z \hat{a}_z^{\dagger} \hat{a}_z + \sum_{j = 1}^N \left[ \frac{\vec{p}_j \cdot \vec{p}_j}{2 m} + \hbar g_x \cos \left( k_x \hat{x}_j \right) \left( \hat{\sigma}_j^+ \hat{a}_x + \mathrm{H.c.} \right) + \frac{\hbar g_z \Omega}{2 \Delta_a} \cos \left( k_z \hat{z}_j \right) \left( \hat{\sigma}_j^- \hat{a}_z + \mathrm{H.c.} \right) \right].
    \end{aligned}
\end{equation}
The effective master equation for the reduced density matrix $\hat{\rho}_{\mathrm{red}} = \sum_j \bra{a}_j \hat{\rho}_1 \ket{a}_j$ therefore becomes
\begin{equation} \label{ReducedMasterEq}
    \pd{\hat{\rho}_{\mathrm{red}}} = - \frac{i}{\hbar} \left[ \tilde{\hat{H}}, \hat{\rho}_{\mathrm{red}} \right] + \hat{\mathcal{D}} \left[ \sqrt{\kappa_x} \hat{a}_x \right] \hat{\rho}_{\mathrm{red}} + \hat{\mathcal{D}} \left[ \sqrt{\kappa_z} \hat{a}_z \right] \hat{\rho}_{\mathrm{red}}.
\end{equation}

\subsection{Elimination of dissipative cavity modes}
Superradiance is expected in the bad-cavity regime, and so we assume $\kappa_x$ and $\kappa_z$ are the largest frequencies in the reduced master equation.
In this parameter regime, the cavity fields can be adiabatically eliminated to find an atomic-only master equation. 
Following the method of Ref.~\cite{Xu2}, we find that the cavity fields become slaved to the atomic degrees of freedom,
\begin{equation} \label{aSlaved}
    \hat{a}_x (t) \approx - \frac{i g_x}{\kappa_x / 2 - i \Delta_x} \sum_{j = 1}^N \cos \left[ k_x \hat{x}_j (t) \right] \hat{\sigma}_j^- (t), \quad \hat{a}_z (t) \approx - \frac{i g_z \Omega}{2 \Delta_a \left( \kappa_z / 2 - i \Delta_z \right)} \sum_{j = 1}^N \cos \left[ k_z \hat{z}_j (t) \right] \hat{\sigma}_j^+ (t),
\end{equation}
where we have ignored Langevin noise. 
We then obtain the master equation for the atomic density matrix $\hat{\rho}_{\mathrm{at}} = \Tr_{x,z}[\hat{\rho}_{\mathrm{red}}]$,
\begin{equation}
    \pd{\hat{\rho}_{\mathrm{at}}} = - \frac{i}{\hbar} \left[ \hat{H}_{\mathrm{at}}, \hat{\rho}_{\mathrm{at}} \right] + \hat{\mathcal{D}} \left[ \sqrt{\frac{\kappa_x g_x^2}{\Delta_x^2 + \kappa_x^2 / 4}} \sum_{j = 1}^N \cos \left( k_x \hat{x}_j \right) \hat{\sigma}_j^- \right] \hat{\rho}_{\mathrm{at}} + \hat{\mathcal{D}} \left[ \sqrt{\frac{\kappa_z g_z^2 \Omega^2}{4 \Delta_a^2 \left( \Delta_z^2 + \kappa_z^2 / 4 \right)}} \sum_{j = 1}^N \cos \left( k_z \hat{z}_j \right) \hat{\sigma}_j^+ \right] \hat{\rho}_{\mathrm{at}},
\end{equation}
with the Hamiltonian,
\begin{equation}
    \hat{H}_{\mathrm{at}} = \sum_{i,j = 1}^N \left[ \frac{\vec{p}_j \cdot \vec{p}_j}{2 m} + \frac{\hbar \Delta_x g_x^2}{\Delta_x^2 + \kappa_x^2 / 4} \cos \left( k_x \hat{x}_i \right) \cos \left( k_x \hat{x}_j \right) \hat{\sigma}_i^+ \hat{\sigma}_j^- + \frac{\hbar \Delta_z g_z^2 \Omega^2}{4 \Delta_a^2 \left( \Delta_z^2 + \kappa_z^2 / 4 \right)} \cos \left( k_z \hat{z}_i \right) \cos \left( k_z \hat{z}_j \right) \hat{\sigma}_i^- \hat{\sigma}_j^+ \right].
\end{equation}
We are mostly concerned with the dissipative terms, so we now assume both cavities are resonant on their respective transitions, $\Delta_x = \Delta_z = 0$.
This gives the master equation
\begin{equation} \label{FullMasterEq}
    \pd{\hat{\rho}_{\mathrm{at}}} = - \frac{i}{\hbar} \left[ \sum_{j = 1}^N \frac{\vec{p}_j \cdot \vec{p}_j}{2 m}, \hat{\rho}_{\mathrm{at}} \right] + \hat{\mathcal{D}} \left[ \sqrt{\Gamma_c} \sum_{j = 1}^N 2 \cos \left( k_x \hat{x}_j \right) \hat{\sigma}_j^- \right] \hat{\rho}_{\mathrm{at}} + \hat{\mathcal{D}} \left[ \sqrt{W} \sum_{j = 1}^N 2 \cos \left( k_z \hat{z}_j \right) \hat{\sigma}_j^+ \right] \hat{\rho}_{\mathrm{at}},
\end{equation}
with the pumping and decay rates
\begin{equation}
    W = \frac{g_z^2 \Omega^2}{4 \Delta_a^2 \kappa_z}, \quad \Gamma_c = \frac{g_x^2}{\kappa_x}.
\end{equation}

\subsection{Reduction to two momentum states} \label{Appendix:MomentumReduction}
We now consider an initial pure state 
\begin{equation}
    \hat{\rho}_{\mathrm{at}} (0) = \op{g, - \frac{\hbar k_x}{2}, - \frac{\hbar k_z}{2}}{g, - \frac{\hbar k_x}{2}, - \frac{\hbar k_z}{2}}^{\otimes N},
\end{equation}
where we have introduced the notation $\ket{i,p_x,p_z}$ with $i \in \{ g, e \}$. 
Therefore, the dynamics under Eq.~\eqref{FullMasterEq} are restricted to the momentum states $\ket{p_x = n \hbar k_x / 2}$ and $\ket{p_z = n' \hbar k_z / 2}$ with integers $n, n' \in \mathbb{Z}$. 
In particular, we are interested in the parameter regime considered in Ref.~\cite{ActuallyJarrodsPaper} in which the energy injected from the cavity fields are less than the energy difference between the $n, n' = 1$ and $n, n' = 3$ states, $N \Gamma_c \ll \hbar^2 k_x^2 / m$ and $N W \ll \hbar^2 k_z^2 / m$. 
Thus, the dynamics are restricted to the momentum states $\ket{p_x = \pm \hbar k_x / 2}$ and $\ket{p_z = \pm \hbar k_z / 2}$.
More specifically, an atom beginning in $\ket{g, - \hbar k_x / 2, - \hbar k_z / 2}$ will only couple to the subspace
\begin{equation}
    \left\{ \ket{g, \frac{\hbar k_x}{2}, \frac{\hbar k_z}{2}}, \ket{g, - \frac{\hbar k_x}{2}, - \frac{\hbar k_z}{2}}, \ket{e, \frac{\hbar k_x}{2}, - \frac{\hbar k_z}{2}}, \ket{e, - \frac{\hbar k_x}{2}, \frac{\hbar k_z}{2}} \right\}
\end{equation}
under the dynamics of Eq.~\eqref{FullMasterEq}.
Since $p_x = p_z$ when in the internal state $\ket{g}$ while $p_x = - p_z$ when in $\ket{e}$, we can drop the $z$-momentum label of the states as it is implied by the internal state and $p_x$. 
As discussed in Sec.~\ref{Sec:Model}, we introduce the label ``$r$'' for $p_x = \hbar k_x / 2$ and ``$l$'' for $p_x = - \hbar k_x / 2$, and so our state space becomes
\begin{equation} \label{MomentumStates}
    \left\{ \ket{g, r}, \ket{g, l}, \ket{e, r}, \ket{e, l} \right\}.
\end{equation}

Defining the Pauli operators for the momentum, $\hat{s}_j^+ = \op{r}{l}_j$ and $\hat{s}_j^x = \hat{s}_j^+ + \hat{s}_j^-$, we can now introduce three collective operators
\begin{equation}
    \hat{J}_{\pm} = \sum_{j = 1}^N \hat{\sigma}_j^{\pm}, \quad \hat{K}_+ = \sum_{j = 1}^N \hat{s}_j^{\pm}, \quad \hat{E}_+ = \sum_{j = 1}^N \hat{\sigma}_j^{\pm} \otimes \hat{s}_j^x,
\end{equation}
which represent the spin degree of freedom, the $x$-momentum degree of freedom, and the entangling process between the two degrees of freedom, respectively.
With this, the master equation Eq.~\eqref{FullMasterEq} for the density matrix in the subspace Eq.~\eqref{MomentumStates}, $\hat{\rho}$,  becomes
\begin{equation}
    \pd{\hat{\rho}} = - \frac{i}{\hbar} \left[ \sum_{j = 1}^N \frac{\vec{p}_j \cdot \vec{p}_j}{2 m}, \hat{\rho} \right] + \hat{\mathcal{D}} \left[ \sqrt{\Gamma_c} \hat{E}_- \right] \hat{\rho} + \hat{\mathcal{D}} \left[ \sqrt{W} \hat{J}_+ \right] \hat{\rho}.
\end{equation}
Noting that the states in Eq.~\eqref{MomentumStates} all have the same kinetic energy, the Hamiltonian becomes a constant multiplied by the identity operator in the subspace and so can be gauge transformed away.
This leaves us with the final form of the atomic master equation,
\begin{equation}
    \pd{\hat{\rho}} = \hat{\mathcal{D}} \left[ \sqrt{\Gamma_c} \hat{E}_- \right] \hat{\rho} + \hat{\mathcal{D}} \left[ \sqrt{W} \hat{J}_+ \right] \hat{\rho}.
\end{equation}

\end{widetext}

\section{Pseudocode of Hybrid Entanglement Entropy Algorithm} \label{Appendix:Algorithm}
Algorithm~\ref{alg:EE} shows pseudocode for the hybrid entanglement entropy algorithm from Ref.~\cite{ActuallyJarrodsPaper3}. Here, we focus on the spin degree of freedom $J$.
For the momentum degree of freedom $K$, the for loop over $j,j',k$ becomes a for loop over $k,k',j$, the for loop over $j,j'$ becomes a for loop over $k,k'$, and the sums $\sum_{j = 0}^{2 l + 1}$ become $\sum_{k = 0}^{2 l + 1}$.
The $l$'s, which label the irreducible representation (``layers'') of the $\mathrm{SU} (2)$ subgroups, must be the same for $J$ and $K$ for the states to have an overall $\mathrm{SU} (4)$ symmetry~\cite{ActuallyJarrodsPaper3}. 
We note that the loop over the ``layers'' $l$ in the algorithm is, in effect, a loop over the irreducible representations of the respective $\mathrm{SU} (2)$ subgroup, and so the multiplicity $d_l$ can be calculated using the method of Young tableaux~\cite{Georgi,Pfeifer,Eichmann}.
The result of this for a given $N$ is given in Eq.~\eqref{d_l}~\cite{Mandel,Xu}.



\begin{algorithm}[t!]
\label{alg:EE}
\SetAlgoLined
Begin with a state $\hat{\rho}$ in $\mathrm{SU} (4)$ bosonic basis\;
Initialize $\hat{J}_-$ and $\hat{K}_-$ in $\mathrm{SU} (4)$ bosonic basis\;
$S_J \leftarrow 0$\;
\For{$l = \frac{N}{2}:-1:0$}{
     $s_v \leftarrow$ Empty $(2 l + 1) \cross (2 l + 1)$ array\;
     \eIf{$l = \frac{N}{2}$}{$\ket{s'} \leftarrow \ket{N,0,0,0}$\;}{Gram-Schmidt orthogonalization for $\ket{s'}$\;}
     \For{$j = 1:2 l + 1$}{
        $\ket{s} \leftarrow \ket{s'}$\;
        $s_a(j,1) = \ket{s}$\;
        \For{$k = 2:2 l + 1$}{
            $\ket{s} \leftarrow \hat{K}_- \ket{s}$\;
            Normalize $\ket{s}$\;
            $s_a(j,k) = \ket{s}$\;
        }
        $\ket{s'} \leftarrow \hat{J}_- \ket{s'}$\;
        Normalize $\ket{s'}$\;
     }
     $P \leftarrow \mathrm{zeros}(2 l + 1, 2 l + 1, 2 l + 1)$\;
     \For{$j,j',k = 1:2 l + 1$}{
        $P(j,j',k) \leftarrow s_a(j,k)^{\dagger} \hat{\rho} s_a(j',k)$\;
     }
     $\mathbf{M}_J^{(l)} \leftarrow \mathrm{zeros}(2 l + 1, 2 l + 1)$\;
     \For{$j,j' = 1:2 l + 1$}{
        $\mathbf{M}_J^{(l)}(j,j') \leftarrow \sum_{k = 0}^{2 l + 1} P(j,j',k) $\;
     }
     $\lambda_J \leftarrow \mathrm{eigenvalues}(\mathbf{M}^{(l)}_J)$\;
     $d_{l} \leftarrow N! (2 l + 1) / [(N / 2 + l + 1)! (N / 2 - l)!]$\;
     $S_J \leftarrow S_J - \sum_{j = 0}^{2 l + 1} \lambda_J(j) \ln(\lambda_J(j) / d_l)$\;
}
\caption{The algebraic entanglement entropy algorithm from Ref.~\cite{ActuallyJarrodsPaper3} for the spin degree of freedom $J$. The algorithm loops over each ``layer'' of states labeled by $l$ from $N/2$ down to $0$ (for $N \in 2 \mathbb{N}$). The starting state for layer $l$ is found by Gram-Schmidt orthogonalization using all previous states whose $\hat{J}_z$ and $\hat{K}_z$ eigenvalues are $l$, but have different eigenvalues of $\hat{J}^2$ (and thus $\hat{K}^2$~\cite{ActuallyJarrodsPaper3}). From this starting state, all other states are reached in the layer by repeated application of $\hat{J}_-$ and $\hat{K}_-$. These states are used to calculate the coefficients of $\hat{\rho}$, and subsequently construct the matrix $\mathbf{M}_J$. Lastly, the eigenvalues of this matrix and the multiplicity $d_{l}$ are used to calculate the algebraic entanglement entropy.}
\end{algorithm}

\section{State Visualization with the Pyramid Representation} \label{Appendix:Pyramid}
In this Appendix, we provide a visualization of the states that are created during the system's evolution using the pyramid structure developed in Ref.~\cite{ActuallyJarrodsPaper3}, and briefly discussed in Appendix~\ref{Appendix:Algorithm}, of stacked $\mathrm{SU} (2) \otimes \mathrm{SU} (2)$ layers. 
Here, one builds the $\ket{l,m_j,m_k}$ basis which are the $s_a(j,k)$ states built in Algorithm~\ref{alg:EE}, where $l$ is the layer's label [which gives the $\hat{J}^2$ and $\hat{K}^2$ eigenvalues $l(l+1)$] while $m_j$ and $m_k$ are the eigenvalues of $\hat{J}_z$ and $\hat{K}_z$, respectively.
Note that $l$ accounts for both the subgroup's Casimir operators' ($\hat{J}^2$ and $\hat{K}^2$) eigenvalues as they must be equal for the $\mathrm{SU} (4)$ state to have overall permutation symmetry~\cite{ActuallyJarrodsPaper3}. 
The layers range from $l \in \{ N/2, N/2 - 1, \ldots, 0 \}$ for even $N$, and since $m_j, m_k \in \{ - l, - l + 1, \ldots, l \}$, each layer can be organized into a $(2 l + 1) \cross (2 l + 1)$ block which gradually decrease in size as $l$ is decreased. 
For the largest layer $l = N/2$, the states are separable in $J$ and $K$, while all states on lower layers $l < N/2$ have intrinsic hybrid entanglement through their $\mathrm{SU} (2)$ degeneracy Eq.~\eqref{d_l}~\cite{ActuallyJarrodsPaper3}.
Stacking the blocks from largest to smallest $l$ leads to a pyramid structure; note that this structure is just a reshaping of a state vector in $\mathrm{SU} (4)$, but the different blocks directly map to the corresponding $\mathrm{SU} (2)$ irreducible representation blocks from Ref.~\cite{Xu} upon tracing out one of the degrees of freedom (which is done in Algorithm~\ref{alg:EE})~\cite{ActuallyJarrodsPaper3}.
We can then make a useful visualization of an $\mathrm{SU} (4)$ state by calculating the overlap with each pyramid state on each block.

\begin{figure*}
    \begin{center}
    \includegraphics[width=\linewidth]{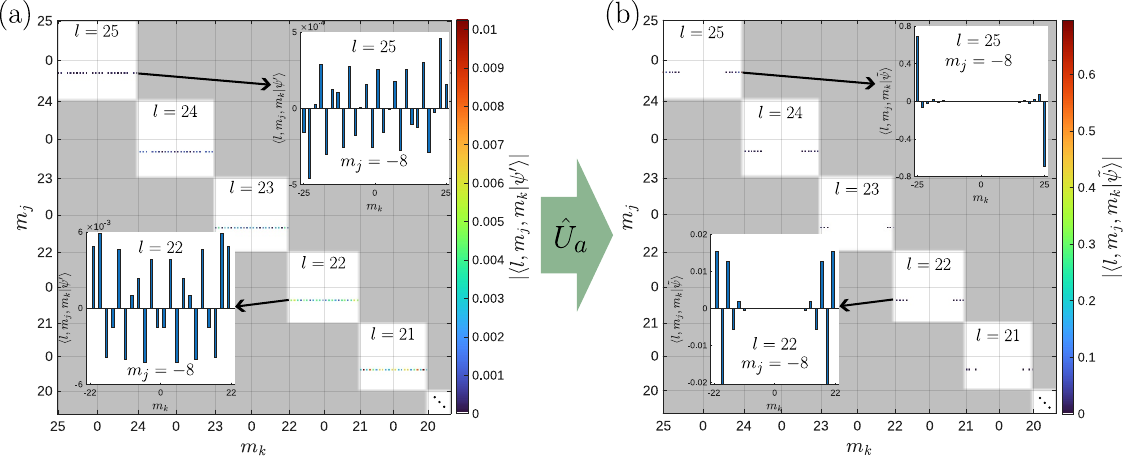}
    \end{center}
    \caption{Visualization of a $\mathrm{SU} (4)$ state using the pyramid representation $\ket{l,m_j,m_k}$. 
    For a Monte-Carlo trajectory that exhibits the behavior of Fig.~\ref{KzFig}(a), we find the state that maximizes the acceleration QFI $\mathcal{F}_a$ after the $\hat{U}_a$ rotation from Eq.~\eqref{RotToKz}.
    We plot the state overlap in the first five layers of the pyramid both (a) before $\ket{\psi'}$ and (b) after $\ket{\tilde{\psi}}$ the rotation $\hat{U}_a$.
    Here, we use $N = 50$ and $W = 0.1 \Gamma_c$.
    The right (left) inset on either plot shows the overlap on layer $l = 25$ ($l = 22$) along the row $m_j = -8$.
    Note that gray areas can not be populated in our system because these states would break the permutation symmetry in $\mathrm{SU} (4)$~\cite{ActuallyJarrodsPaper3}.}
    \label{PyramidOverlap}
\end{figure*}
As an example of this state visualization, we examine the pure state $\ket{\tilde{\psi}}$ that maximizes the QFI for acceleration $\mathcal{F}_a$ after the rotation Eq.~\eqref{RotToKz} during a Monte-Carlo wave function simulation from Sec.~\ref{Sec:Metrology}.
For a trajectory that exhibits the behavior of Fig.~\ref{KzFig}(a) with $W = 0.1 \Gamma_c$ and $N = 50$, we plot the highest $\mathcal{F}_a$ state both (a) before $\ket{\psi'}$ and (b) after $\ket{\tilde{\psi}}$ the rotation $\hat{U}_a$ from Eq.~\eqref{RotToKz} in Fig.~\ref{PyramidOverlap}. 
This state $\ket{\psi'}$ has a maximum QFI of $\lambda_{\mathrm{max}} = 0.547 N^2$ with the optimal generator [Eq.~\eqref{OptGenerator}] $\hat{\mathcal{G}} = 0.457 \hat{\mathcal{U}}_x - 0.889 \hat{\mathcal{V}}_x$ [this state also has the largest $\lambda_{\mathrm{max}}$ throughout the trajectory], while the rotated state $\ket{\tilde{\psi}}$ has an acceleration QFI of $\mathcal{F}_a = 0.496 N^2$.
Figure~\ref{PyramidOverlap} shows that the state population is distributed over the different layers but always occupies a single row $m_j = -8$ in every layer. 
This is because the state begins in a $\hat{J}_z$ eigenstate [Eq.~\eqref{InitialState}] while the jump operators $\hat{J}_+$ and $\hat{E}_-$ belong to root spaces that raise or lower the $\hat{J}_z$ weight by $\pm 1$ as $[\hat{J}_z, \hat{J}_{\pm}] = \pm \hat{J}_{\pm}$ and $[\hat{J}_z, \hat{E}_{\pm}] = \pm \hat{E}_{\pm}$ (and thus eigenstates in the adjoint representation)~\cite{Georgi}. 
Consequently, throughout the Monte-Carlo simulation the state is always along a single $m_j$ which is constant during the evolution under $\hat{H}_{\mathrm{NH}}$ and then changes by $\pm 1$ when jumps occur. 
$\hat{J}_+$ jumps stay in the same layer because it commutes with $\hat{J}^2$ and $\hat{K}^2$ while $\hat{E}_-$ jumps can switch layers because it does not.
Moreover, the state population is either symmetric or anti-symmetric about $m_k = 0$ at steady-state. 
To demonstrate this, we show the overlap along the $m_j = -8$ row of the $l = 25$ ($l = 22$) layer in the right (left) inset of both subplots of Fig.~\ref{PyramidOverlap}.
Upon the rotation $\hat{U}_a$ from Eq.~\eqref{RotToKz}, the state goes from having population in $\ket{m_k}$ states throughout the entire $-l$ to $l$ range to having population primarily near $m_k \approx -l$ and $m_k \approx l$ with little population near $m_k \approx 0$, i.e., population on each $l$ shell of the $K$ collective Bloch sphere is concentrated near the shell's north and south pole with little population near the equator. 
This means that the rotated state has NOON-like, or more generally GHZ-like, properties where population in the north and south pole interfere in a manner where narrow fringes form around the equator~\cite{Degen,Pezze}.
This helps explain why the generated states are so sensitive to $\hat{K}_z$ rotations leading to a $\mathcal{F}_a$ that is half the Heisenberg limit. 

\begin{figure}
    \centerline{\includegraphics[width=\linewidth]{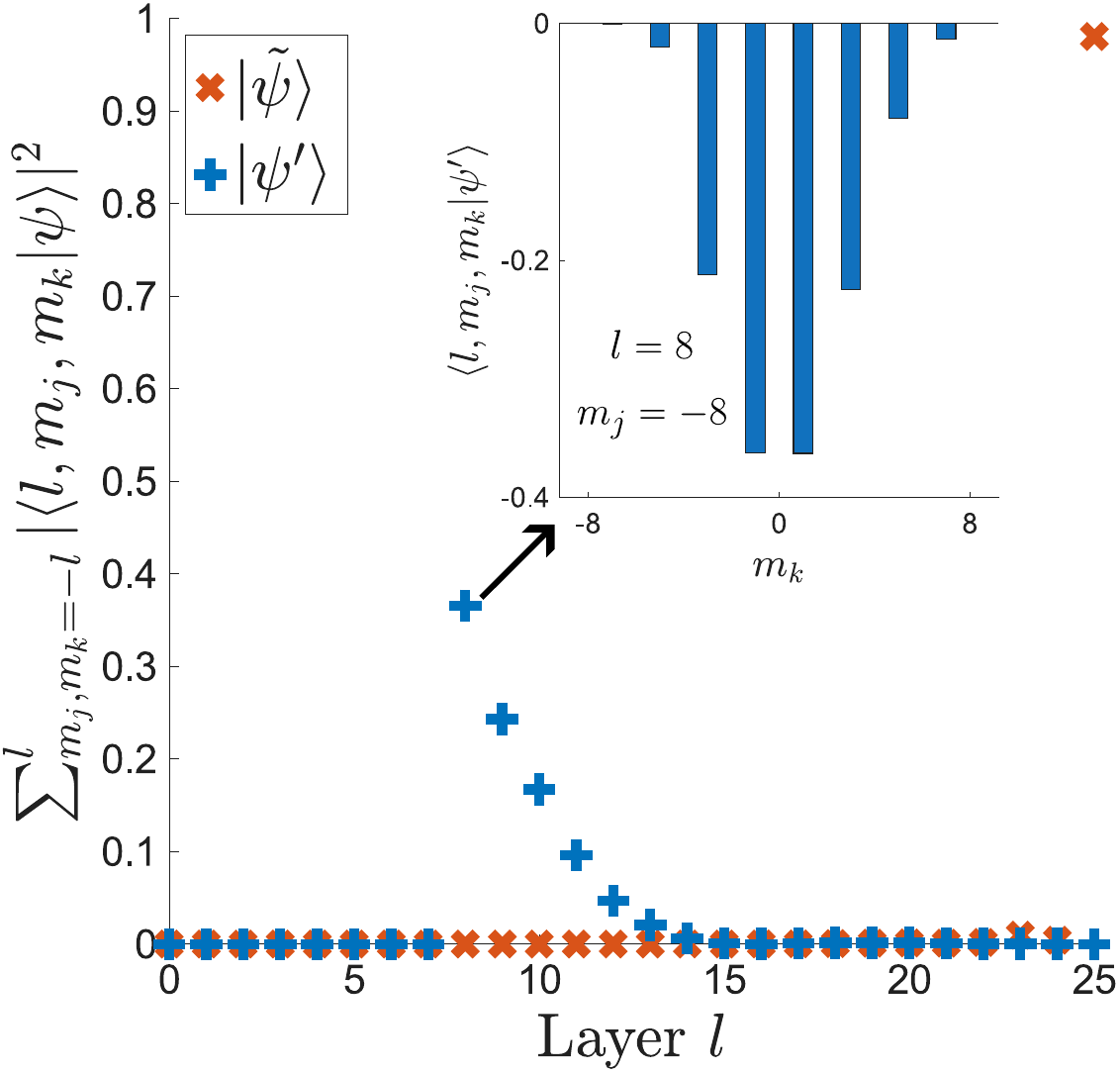}}
    \caption{The populations in each layer for the non-rotated $\ket{\psi'}$ and rotated $\ket{\tilde{\psi}}$ states displayed in Fig.~\ref{PyramidOverlap}.
    The inset shows the overlap of $\ket{\psi'}$ in its most populated layer $l = 8$ (again along the row $m_j = -8$).}
    \label{LayerPops}
\end{figure}
It is important to note that the $\hat{U}_a$ rotation commutes with $\hat{J}_z$ such that the $m_j$ value does not change, but $\hat{U}_a$ does not commute with the $J$ and $K$ subgroups' Casimir operators, and so the rotation can change the population in each layer $l$. 
To demonstrate this, we plot the population in each $l$ layer for the non-rotated (blue) and rotated (orange) states in Fig.~\ref{LayerPops}.
For both states, we find that the layers with $l < 8$ are not populated because they do not have a $m_j = - 8$ state.
The non-rotated state's population is concentrated near $8 \leq l \lesssim 14$ and we show the population along the $m_j = - 8$ row for the highest populated layer $l = 8$ in the inset of Fig.~\ref{LayerPops}.
Meanwhile, the rotated state's population is concentrated in the largest layers $22 \lesssim l \leq 25$ with the base layer $l = 25$ being by far the most populated layer. 
It would be interesting to explore whether this rotation of population to be almost entirely in the $l = N / 2$ layer is a general property of states in trajectories like Fig.~\ref{KzFig}(a) and whether this can help explain why certain runs have this behavior as opposed to trajectories like Fig.~\ref{KzFig}(b). 
Further investigation on the states in Monte-Carlo simulations, especially comparing trajectories exhibiting the behavior of Fig.~\ref{KzFig}(a) compared to Fig.~\ref{KzFig}(b), is left for future work.

\section{Mean-Field Dynamics} \label{Appendix:MF}
In this Appendix, we discuss the mean-field description of the spin-momentum superradiance model.
The central assumption of the mean-field approximation is that the many-body density matrix factorizes into a product of single-particle density matrices,
\begin{equation}
	\hat{\rho}=\bigotimes_{j=1}^{N}\hat{\rho}_j,
\end{equation}
where $\hat{\rho}_j$ denotes the density matrix of atom $j$. Furthermore, we assume that all single-particle density matrices are identical. This follows from the permutation symmetry of the model together with the assumption of a permutationally symmetric initial state. We therefore write
\begin{equation}
	\hat{\rho}_j=\hat{\rho}_{\mathrm{mf}}.
\end{equation}

The mean-field dynamics are obtained from the full master equation~\eqref{AtomicMasterEq} by tracing out all particles except one, which we take to be the first particle. This yields
\begin{equation} \label{eq:mfmaster}
	\frac{\partial \hat{\rho}_{\mathrm{mf}}}{\partial t}
	=
	- \frac{i}{\hbar}
	\left[
	\hat{H}_{\mathrm{mf}},
	\hat{\rho}_{\mathrm{mf}}
	\right],
\end{equation}
where the self-consistent mean-field Hamiltonian is given by
\begin{equation}
	\hat{H}_{\mathrm{mf}}
	=
	-iNW\left(
	j^{*}\hat{j}
	-
	\hat{j}^{\dagger}j
	\right)
	+
	iN\Gamma_c\left(
	e^{*}\hat{e}
	-
	\hat{e}^{\dagger}e
	\right),
\end{equation}
with the mean fields
\begin{equation}
    \begin{aligned}
e &=
	\mathrm{Tr}_1
	\!\left[
	\hat{e}\hat{\rho}_{\mathrm{mf}}
	\right],\\
j &=
	\mathrm{Tr}_1
	\!\left[
	\hat{j}\hat{\rho}_{\mathrm{mf}}
	\right].
    \end{aligned}
\end{equation}
Here, $\mathrm{Tr}_1$ denotes the trace over the single-particle Hilbert space, and the relevant single-particle operators are
\begin{equation}
    \begin{aligned}
\hat{e} &= \ket{g,l}\bra{e,r}+\ket{g,r}\bra{e,l}\,\\
\hat{j} &=\ket{g,l}\bra{e,l}+\ket{g,r}\bra{e,r}.
    \end{aligned}
\end{equation}
In deriving Eq.~\eqref{eq:mfmaster}, we have neglected terms of order $1/N$.

Throughout the following analysis, the mean-field density matrix is initialized according to Eq.~\eqref{InitialState},
\begin{equation} \label{eq:mfini}
	\hat{\rho}_{\mathrm{mf}}(0)
	=
	\ket{g,l}\bra{g,l}.
\end{equation}
Since Eq.~\eqref{eq:mfmaster} describes purely unitary evolution under the effective Hamiltonian $\hat{H}_{\mathrm{mf}}$, a pure initial state remains pure at all times. Consequently, the single-particle density matrix $\hat{\rho}_{\mathrm{mf}}$ can be represented by a state vector throughout the evolution. Collective many-body effects enter the mean-field description solely through the self-consistency condition, which renders Eq.~\eqref{eq:mfmaster} nonlinear in $\hat{\rho}_{\mathrm{mf}}$.

First, the initial state in Eq.~\eqref{eq:mfini} is dynamically unstable for any $W>0$, independent of the ratio $W/\Gamma_c$. This may appear surprising, as one might expect that for $W < \Gamma_c$ the system should rapidly decay back to the ground state. The origin of this instability lies in the structure of the accessible state space, which allows the system to access states exhibiting correlations between the internal states $\ket{g}, \ket{e}$ and external degrees of freedom $\ket{l}, \ket{r}$. These correlated states can have a small $j$ coherence and therefore exhibit slow collective decay.

\begin{figure}
	\centerline{\includegraphics[width=\linewidth]{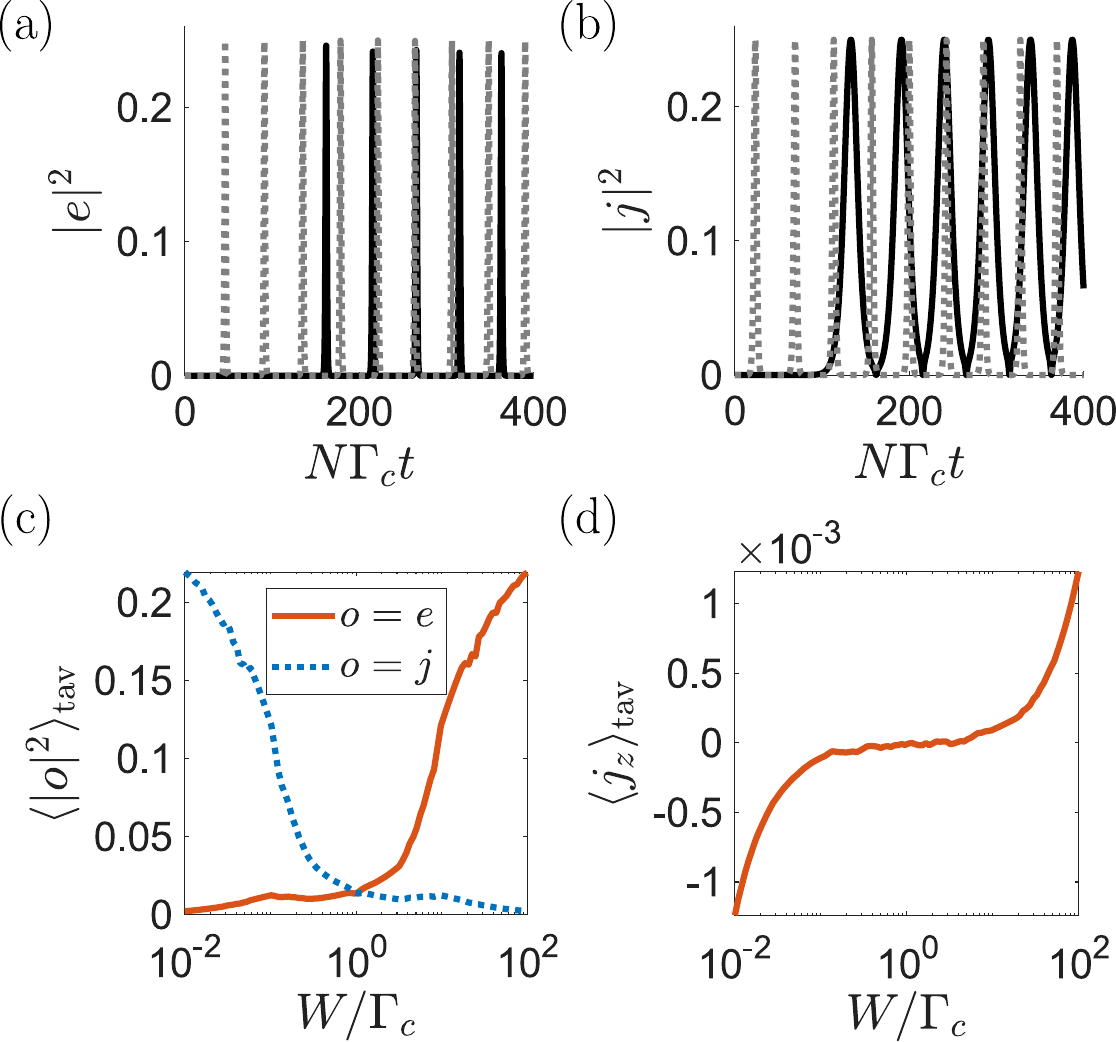}}
	\caption{Coherences (a) $|e|^2$ and (b) $|j|^2$ obtained by evolving Eq.~\eqref{eq:mfmaster}, for $W=0.1\Gamma_c$ (black solid) and $W=0.5\Gamma_c$ (gray dotted). 
    (c) Time-averaged coherences $\langle |e|^2\rangle_{\mathrm{tav}}$ (orange solid) and $\langle |j|^2\rangle_{\mathrm{tav}}$ (blue dotted) obtained by evolving Eq.~\eqref{eq:mfmaster} for times $T=6000/\min(NW,N\Gamma_c)$ and then using Eqs.~\eqref{etav} and~\eqref{jtav}. 
    (d) Time-averaged inversion $\langle j_z \rangle_{\mathrm{tav}}$ obtained in the same way as in (c) using Eq.~\eqref{jztav}.}
    \label{Meanfield}
\end{figure}
In Fig.~\ref{Meanfield}(a) we show the values of $|e|^2$ and (b) $|j|^2$ as a function of time for $W=0.1\Gamma_c$ (black solid) and $W=0.5\Gamma_c$ (gray dotted). For both values of $W$ we see oscillatory dynamcis: the shape of $|e|^2$ and $|j|^2$ appears in bursts while the burst of $|j|^2$ are slightly longer for $W=0.1\Gamma_c$. This analysis shows that the mean-field dynamics does not find a true stationary state. Instead it is dominated by fast dynamics and large oscillations. As the cavity fields are locked to the $\hat{E}$ and $\hat{J}$ dipoles and $e$ and $j$ are the mean-field equivalent of these, the mean-field model would predict quasi infinitely long dynamics of bursts. 

To obtain a stationary characterization of the mean-field dynamics despite the absence of relaxation, we consider the time-averaged observables
\begin{align}
		\langle |e|^2 \rangle_{\mathrm{tav}}
	&=
	\frac{1}{T}\int_0^T dt\, |e(t)|^2,\label{etav}
	\\
    \langle |j|^2 \rangle_{\mathrm{tav}}
	&=
	\frac{1}{T}\int_0^T dt\, |j(t)|^2,\label{jtav}
	\\
	\langle j_z \rangle_{\mathrm{tav}}
	&=
	\frac{1}{T}\int_0^T dt\,
	\mathrm{Tr}_1
	\!\left[
	(\ket{e}\bra{e}-\ket{g}\bra{g})
	\hat{\rho}_{\mathrm{mf}}(t)
	\right].\label{jztav}
\end{align}
The resulting values of $\langle |e|^2 \rangle_{\mathrm{tav}}$ and $\langle |j|^2 \rangle_{\mathrm{tav}}$, obtained after an evolution time
$T=6000/\min(NW,N\Gamma_c)$, are shown as the orange and blue curves in Fig.~\ref{Meanfield}.
For $W<\Gamma_c$, the coherence is predominantly contained in $\langle |j|^2 \rangle_{\mathrm{tav}}$, while $\langle |e|^2 \rangle_{\mathrm{tav}}$ remains small. The opposite behavior is found for $W>\Gamma_c$. Thus, the mean-field description reproduces the qualitative behavior observed in the full many-body simulations. In particular, Fig.~\ref{Meanfield} may be compared with Fig.~\ref{LightOutput}(a),(b), where the dominant coherence is likewise found in $\langle \hat{J}_-\hat{J}_+ \rangle$ for $W<\Gamma_c$ and in $\langle \hat{E}_+\hat{E}_- \rangle$ for $W>\Gamma_c$.

Quantitatively, however, substantial differences emerge. Notably, the plateau-like behavior of $\langle \hat{E}_+\hat{E}_- \rangle$ for $W>\Gamma_c$ and of $\langle \hat{J}_-\hat{J}_+ \rangle$ for $W<\Gamma_c$, clearly visible in Fig.~\ref{LightOutput}, is absent in the mean-field results. These discrepancies are further emphasized by the inversion $\langle\hat{J}_z\rangle$. Comparing the inset of Fig.~\ref{LightOutput}(b) with Fig.~\ref{Meanfield}(d), we find that mean field predicts both significantly smaller inversion values and a qualitatively different dependence on $W/\Gamma_c$. Whereas the many-body inversion saturates for both small and large values of $W/\Gamma_c$, the time-averaged mean-field inversion increases monotonically throughout the displayed parameter range.

We attribute these discrepancies to the neglect of correlations in the mean-field approximation. By construction, the factorized ansatz excludes both quantum and classical correlations. Moreover, the mean-field dynamics remain purely coherent and therefore lack both decoherence and genuine relaxation mechanisms. As a consequence, while the mean-field description might capture some aspects of the short-time dynamics and reproduces the qualitative redistribution of coherence across the transition, it fails to accurately describe the stationary-state properties of the system.

\clearpage

\bibliography{references.bib}

\end{document}